\documentclass[11pt,twoside]{article}

\newcommand{\sindex}[1]{\index{#1}}
\newcommand{\oindex}[1]{\index{#1}}

\newcommand{\sxindex}[1]{\index{#1}}

\usepackage{asp2004}
\usepackage{epsf}
\usepackage{psfig}
\usepackage{lscape}
\usepackage{multirow}
\usepackage{bigdelim}
\usepackage{graphicx}
\usepackage{amsmath}
\usepackage{relsize}
\usepackage{makeidx}
\usepackage[hyperindex]{hyperref}
\hypersetup{pdftex=true, colorlinks=true, breaklinks=true, linkcolor=blue, menucolor=black, pagecolor=blue, urlcolor=blue}
\usepackage{breakurl}

\makeindex

\begin{document}

\markboth{Freudling et al.}{FORS Photometry}    
\pagestyle{myheadings}
\setcounter{equation}{0}
\setcounter{figure}{0}
\setcounter{footnote}{0}
\setcounter{section}{0}
\setcounter{table}{0}

\title{Photometry with FORS at the ESO VLT}    
\author{W. Freudling, M. Romaniello, F. Patat, P. M{\o}ller,
E. Jehin,  K. O'Brien} 
\affil{European Southern Observatory\\Karl-Schwarzschild-Strasse 2\\85748 Garching bei M\"unchen,Germany}

\begin{abstract} 

ESO's two {\bf FO}cal {\bf Re}ducer and low dispersion {\bf S}pectrographs
(FORS)  are the primary imaging cameras for the VLT. Since they are not
direct-imaging cameras, the accuracy of photometry which can routinely be
obtained is limited by significant sky concentration and other effects.

Photometric standard observations are routinely obtained by ESO, and nightly
zero points are computed  mainly for the purpose of monitoring the instrument
performance.  The accuracy of these zero points is about 10\%. 

Recently, we have started a program to investigate, if and  how percent-level
absolute photometric accuracy with FORS can be achieved.  The main results of
this project are presented in this paper.  We first discuss the quality of the
flatfields and how it can be improved.  We then use data with improved
flat-fielding to investigate the usefulness of \sindex{Stetson standard fields}
for FORS calibration and the accuracy which can be achieved. 

The main findings of the FORS Absolute Photometry Project program are as
follows.  There are significant differences between the sky flats and the true
photometric response of the instrument which partially  depend on the rotator
angle. A second order correction to the sky flat significantly improves the
relative photometry within the field. Percent level photometric accuracy can be
achieved with FORS1.  To achieve this accuracy, observers need to invest some
of the assigned science time for imaging of  photometric standard fields in
addition to the routine nightly photometric calibration. 

\end{abstract}

\section{Introduction}

ESO operates two version of the  {\bf FO}cal {\bf Re}ducer and low dispersion {\bf S}pectrograph
\citep[FORS,][]{fors}, FORS 1 and FORS 2.  Routine nightly
photometric calibration observations for the FORS and other ESO imaging cameras  aim
to provide photometric accuracy of about 5 to 10\%. The primary purpose
of these observations is to monitor instrument performance. In addition, they
often are used to calibrate science observations which do not require highly
accurate photometry.  The ESO  {\em FORS Absolute Photometry Project (FAP)}
recently used FORS1 to investigate the accuracy of photometry with that camera
with the goal to establish procedures and advice observers with percent level
photometric calibration needs.  The specific goal of {FAP} was to
demonstrate the feasibility of 3\% photometry.  This report describes the
methods and results  of {\em FAP}.

\section{FORS Flats}

The quality of the flatfields determines to a large extend the accuracy of
relative photometry.  For FORS, twilight sky flats are used almost
exclusively.  \sindex{$UBVRI$} twilight flats are routinely taken at the start
and end of the night, usually in groups of 4 frames.  One area of concern in
any \sindex{flat-fielding procedure} is the presence of  large scale features
in the flatfields which do not correspond to variations in the sensitivity as a function of
position on the detector.  If for example illumination gradients are present,
they  will be propagated into the science images and the resulting photometry
will be affected by position-dependent systematic errors. 

Such gradients in flatfields are often introduced by the illumination source,
in the case of FORS flats  the twilight sky. Gradients or other flatfield
artifacts can also be introduced by the instrument, e.g. through scattered
light. Both issues are examined in this section.\sxindex{gradients in
flatfields}\sxindex{sky gradients}

\subsection{Sky Gradients}\label{sec:flat}

During twilight, the sky is known to show illumination gradients, which change
with time and the position of the Sun relative to the pointing of the
telescope.   Under conditions which are typical for FORS sky flats, the
measured gradients can range from 2 to 5\% per degree \citep[][]{chromey}.
For the field of view (FOV) of FORS (6$^\prime$.8$\times$6$^\prime$.8), this
translates into natural gradients that range from 0.2 to 0.5\%.  On these small
spatial scales, the illumination pattern is expected to be  well approximated
by an inclined plane, whose maximum gradient direction and intensity changes
with the position of the Sun relative to the  imaged sky.  In principle, such
sky gradients can be removed from individual flats before stacking them
\citep{chromey}.

\begin{figure}[!ht]
\centerline{\includegraphics[width=10cm,angle=0,clip]{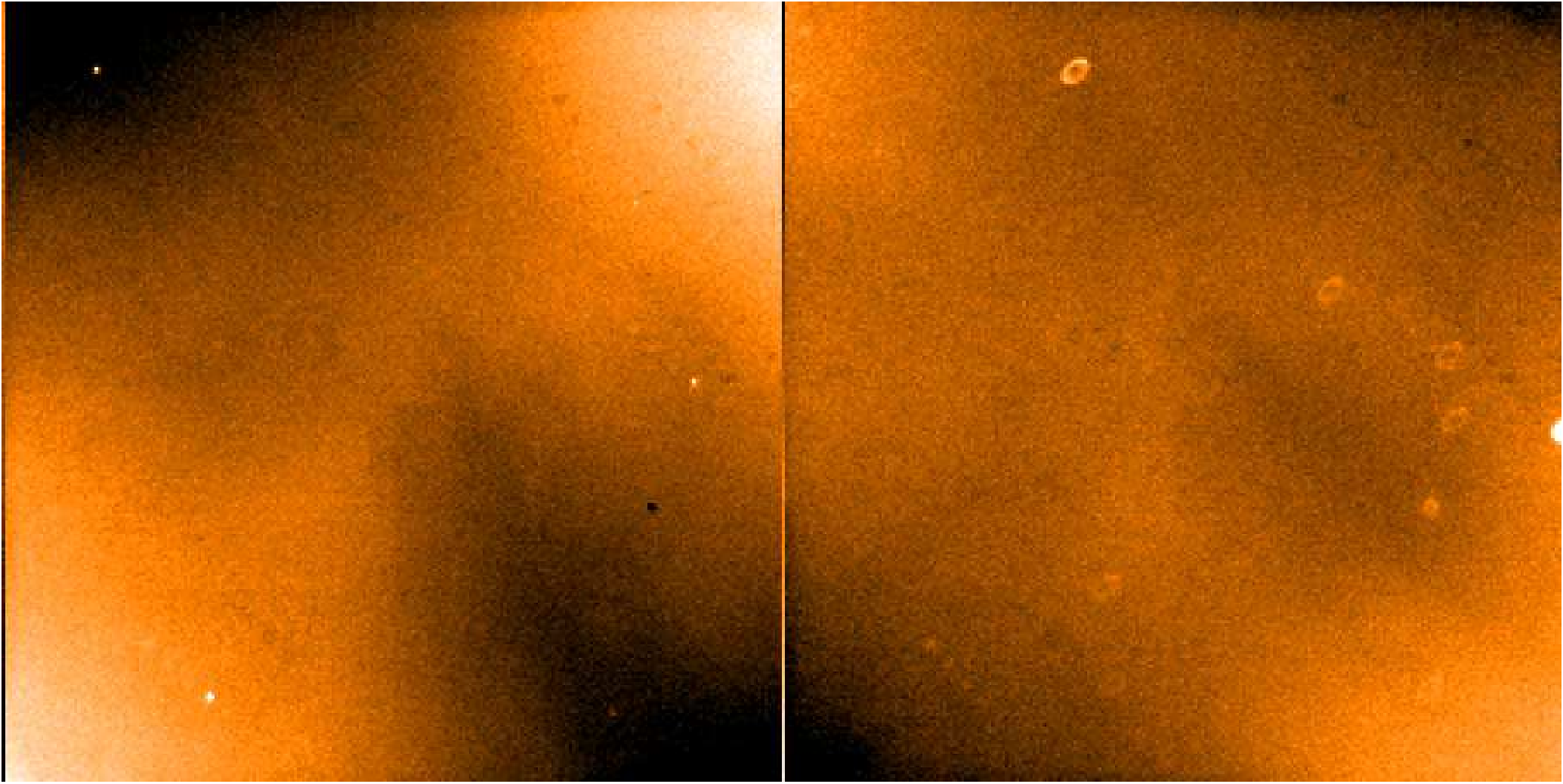}}
\caption[Comparison between two $R$ flats.]{\label{fig:grad} Comparison between
two $R$ flats taken on July 14, 2005 on 10:59:19.530 (left) and 22:43:32.037
(right). The images are flatfields divided by the mean of all flatfields for
that night. The
intensity scale range is 3\%.} \end{figure}

The  structure in observed FORS flatfields  reaches peak-to-peak values of more
than 3\%, i.e. an order of magnitude higher than typical twilight sky
gradients. The structure in flat fields taken in consecutive or even the same
night sometimes differs substantially. This is illustrated in
Fig.~\ref{fig:grad}, were we compare two FORS1 $R$ band flats taken about 12
hours apart.  These changing  structures, which  dominate over the natural
twilight sky gradients, make it difficult to judge whether such gradients are
present in any given observed flatfield.

\subsection{Instrumental features}\label{sec:rotator}

In order to investigate  the structure and amplitude of FORS1 intrinsic
flatfield more systematically, we  have retrieved from the ESO Archive all
\sindex{twilight sky-flats} obtained between January 1, 2005 and September 30,
2005 in the $UBVRI$ passbands .  The total number of images taken with standard
resolution, 4-port read-out and high gain setting is 1083  ($U$: 148, $B$:
208, $V$: 226, $R$: 261, $I$: 240).\sxindex{bias correction}

\begin{figure}[!h]
\centerline{\includegraphics[width=10.2cm,angle=0,clip]{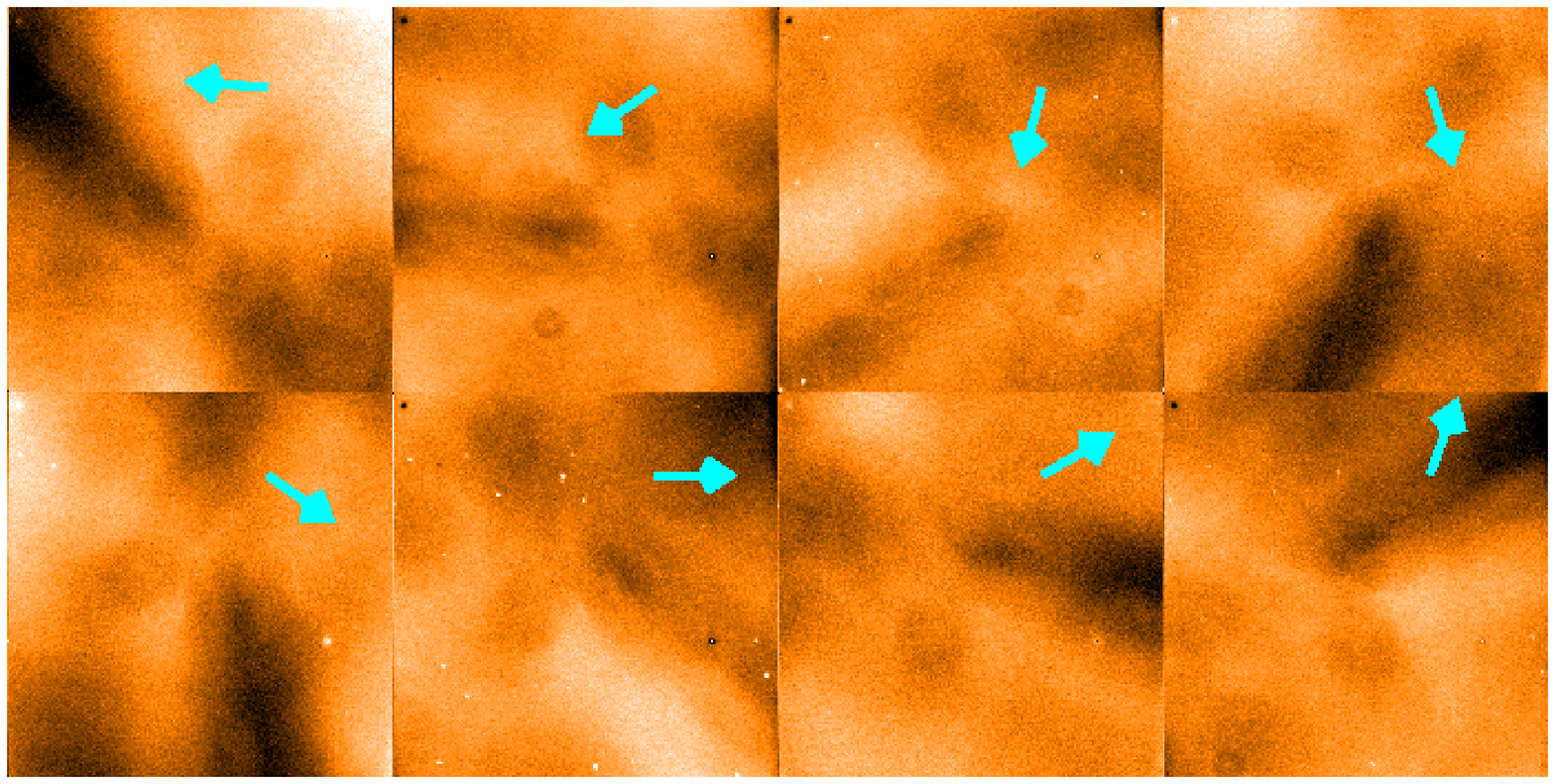}}
\vspace{1mm}
\centerline{\includegraphics[height=5.cm,angle=0,clip]{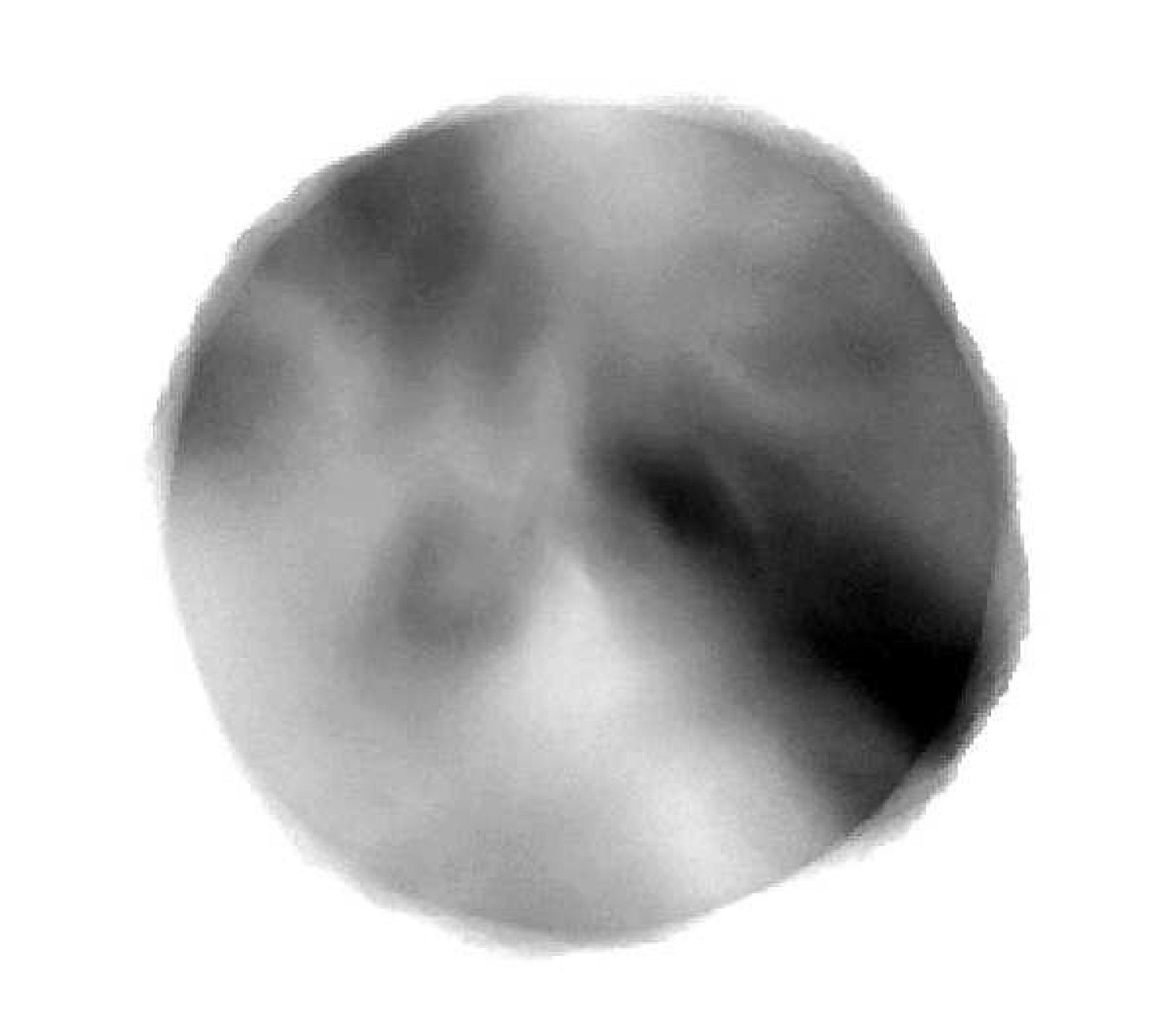}}
\caption[Twilight sky flat in the $R$ passband.]{\label{fig:rflat} {\sl Top}: A
sequence of $B$ FORS1 sky-flats. The rotator adapter angles ($-105^\circ,
-73^\circ, -35^\circ, 0^\circ, +30^\circ$, and  +$70^\circ$) are indicated by a
blue arrow in the upper right corner.  {\sl Bottom:} Stack of rotated $B$-band
sky flats.  Each image was rotated by the negative of the adaptor rotator angle
} \end{figure}

Each individual image  was then bias-corrected using the pre-scan region only.
We then computed the mean of the flatfields for each filter, and divided each
individual frame by this mean. This removed  the stable part of the flatfields
such as the difference between the four quadrants and other large scale
features.  Finally, to allow quick visual inspection, we produced movies
where each frame is an individual sky flat.  Inspection of such movies revealed
that the structure in the flatfields consists of a temporally constant pattern
superimposed on large scale fluctuations which  rapidly change in time.  The
contrast of the constant pattern is higher in  bluer bands. We also found a
correlation of some of the patterns with the adaptor rotator angle.  FORS1 is
mounted on an  adapter rotator which compensates for the sky field rotation
inherent to the VLT alt-azimuth mounting.  Part of the structure in the flat
field seems to  rotate rigidly with the angle of the rotator.  This is
illustrated in  in Fig.~\ref{fig:rflat}. This pattern in the flatfield must be
external to FORS1 and might  be due to reflections and/or asymmetric vignetting
within the telescope or the adapter itself.

We extracted a high signal-to-noise version of the rotating structure in the
following manner. First, we counter-rotated all $B$ images by an amount equal
to the rotator angle reported in their FITS header. We then computed the median
of the rotated flatfields.  The resulting median image is shown in
Fig.~\ref{fig:rflat}.  If there were no correlation between the structure in
the flatfields and the rotator angle, then the structure of the individual
flatfields should average out and the median would be smooth and flat.  Instead,
the opposite can be seen can be seen in Fig.~\ref{fig:rflat}. A finger-like
pattern, which is already visible in the individual flats shown in
Fig.~\ref{fig:rflat}, stands out with increased signal-to-noise.  It
is therefore  clear that this feature rotates. 
The peak-to-peak amplitude of the pattern is about 1\%. Inspection of
individual images in the stack shows that the amplitude varies substantially
among the individual flatfields.

\begin{figure}[!h]
\centerline{\includegraphics[height=5.cm,angle=0,clip]{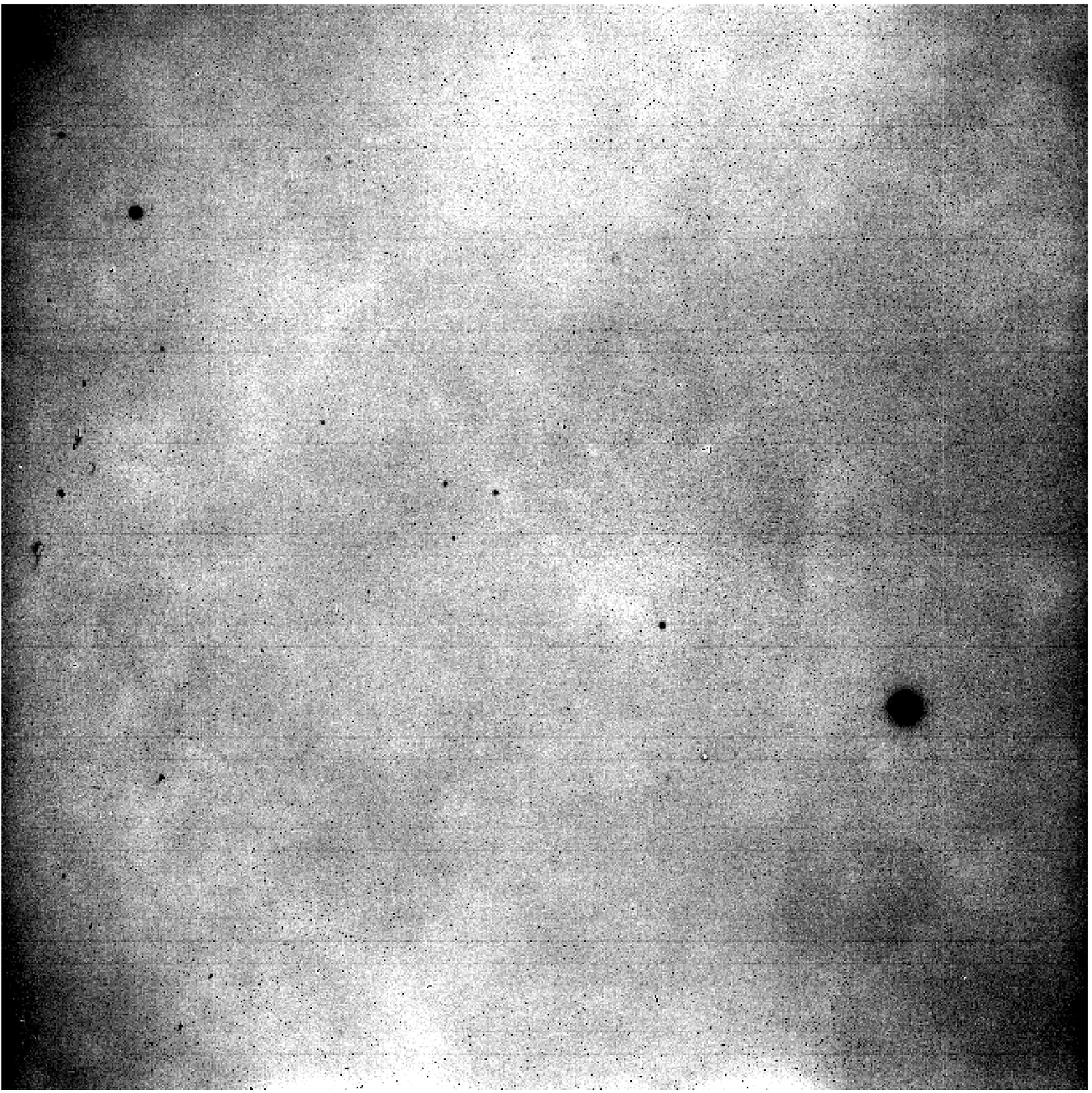}\hspace{16mm}{\includegraphics[height=5.0cm,angle=0,clip]{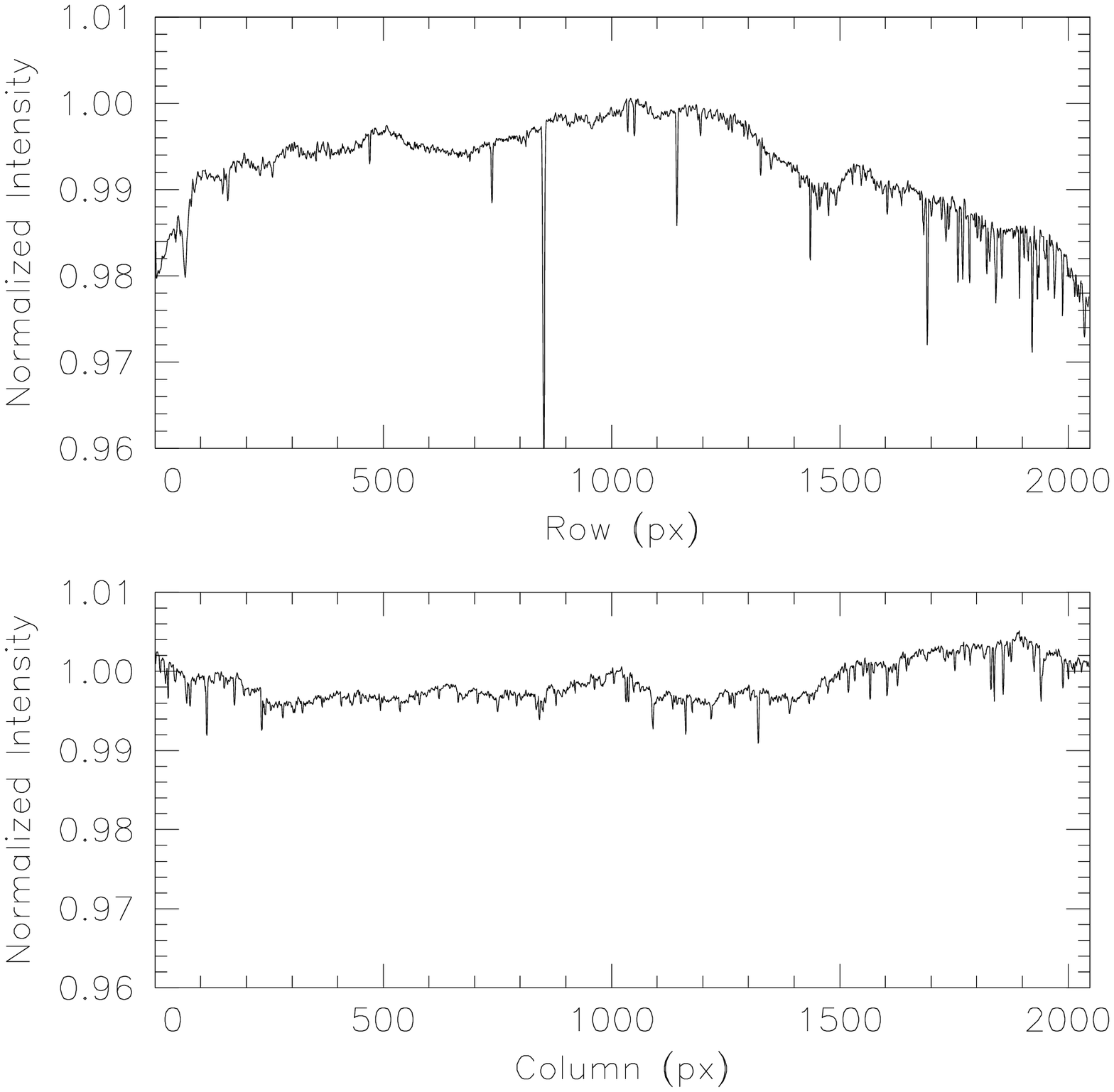}}}
\caption{\label{fig:kiloflat}Co-addition of a selected sample of 240 $R$ sky
flats.  {\sl Left}:   Image of the combined flatfield.  The intensity scale
range is 3\%.  {\sl Right}: Traces along the central column (upper panel) and
central row (lower panel).} \end{figure}

The existence of such a pattern imposes a limit on the accuracy of relative
photometry reachable with FORS.  If the feature is due to a rotation of the
sensitivity pattern imprinted on all science frames, one would need to
carefully match the rotator angle of the flatfield to those of each science
frame.  If,  however,  this feature is  an additive component to the flat,
removing it from the flatfield would improve the photometric accuracy of all
flatfielded science data.  One possible procedure to remove the rotating
feature is  to combine a large number of flatfield images taken with different
rotator angles. If all angles are represented with the same weight, any
rotating structure should smooth out.  We tested this procedure with the $R$
band flatfields. In each rotator angle bin of $10^\circ$, we selected the same
number of flatfields.  We then computed the median of those flats.  The result
is shown in Fig.~\ref{fig:kiloflat}. 
The remaining
structure in this combined flat is much more rotationally symmetric than the one in  individual
frames.  There appears to be a  small central light concentration.

Both, the rotating features shown in Fig.~\ref{fig:rflat} and the illumination
pattern shown in Fig.~\ref{fig:kiloflat} appear to be stable in  time, at least
within the 9 month explored by this investigation. However, it should be
emphasised that not all the variations in the flatfields can be described by a
simple rotation. For example, the difference between the two flatfields shown
in Fig.~\ref{fig:grad} seems to be similar to a $90^\circ$ rotation, but the
adapter rotator angle changed by only about $25^\circ$. In
Sec.~\ref{sec:secondorder} we will therefore investigate how FORS1 flat fields
can be further improved.

\subsection{Impact on Photometry}

The key finding of this Section is that relative gradients in individual
twilight flats as routinely obtained each night differ from each other by as
much as 5\%.  If such flatfields are applied to science data, the relative
photometric accuracy is limited to about 5\%. Even when controlled for rotator
angle, flatfields differ from each other by an amount which questions the goal
of the current project, i.e.  relative photometric accuracy of better than 3\%.
A key question is whether these fluctuations reflect true differences in the
end-to-end throughput of FORS1. In that case, relative and therefore absolute
accuracy at the percent level simply cannot be obtained with FORS1. An
alternative explanation is that the flatfields are flawed and do not represent
the throughput of FORS1. In that case, the task is to find the true flatfield
which should be applied to data so that photometric zero points are constant
over the whole detector. In the following sections, we will use data from the
{\em FAP} programme to test the quality of the flatfields constructed in this
section, and compare it to the regular ``master flats" produced by combining
the routinely taken twilight flats for that night.

\section{FAP Data}

\subsection{Observations}\label{sec:obs}

Obtaining 3\% photometric accuracy requires: 1) relative photometric
calibration within each field; 2) absolute calibration of the extinction
relation with slope and zero point; and 3) calibration of colour terms.  Methods
and data to obtain accurate relative photometry within the FORS1 field have
been presented by \citeauthor{fzwg} in  report~II of the FSSWG project
\citep[][hereafter FWII]{fzwg}.  For the current project, we aimed at an
independent assessment of the relative photometric calibration to investigate
whether the FWII results can be reproduced. 

\begin{figure}[!h] 
\centerline{\includegraphics[width=7cm,angle=0,clip]{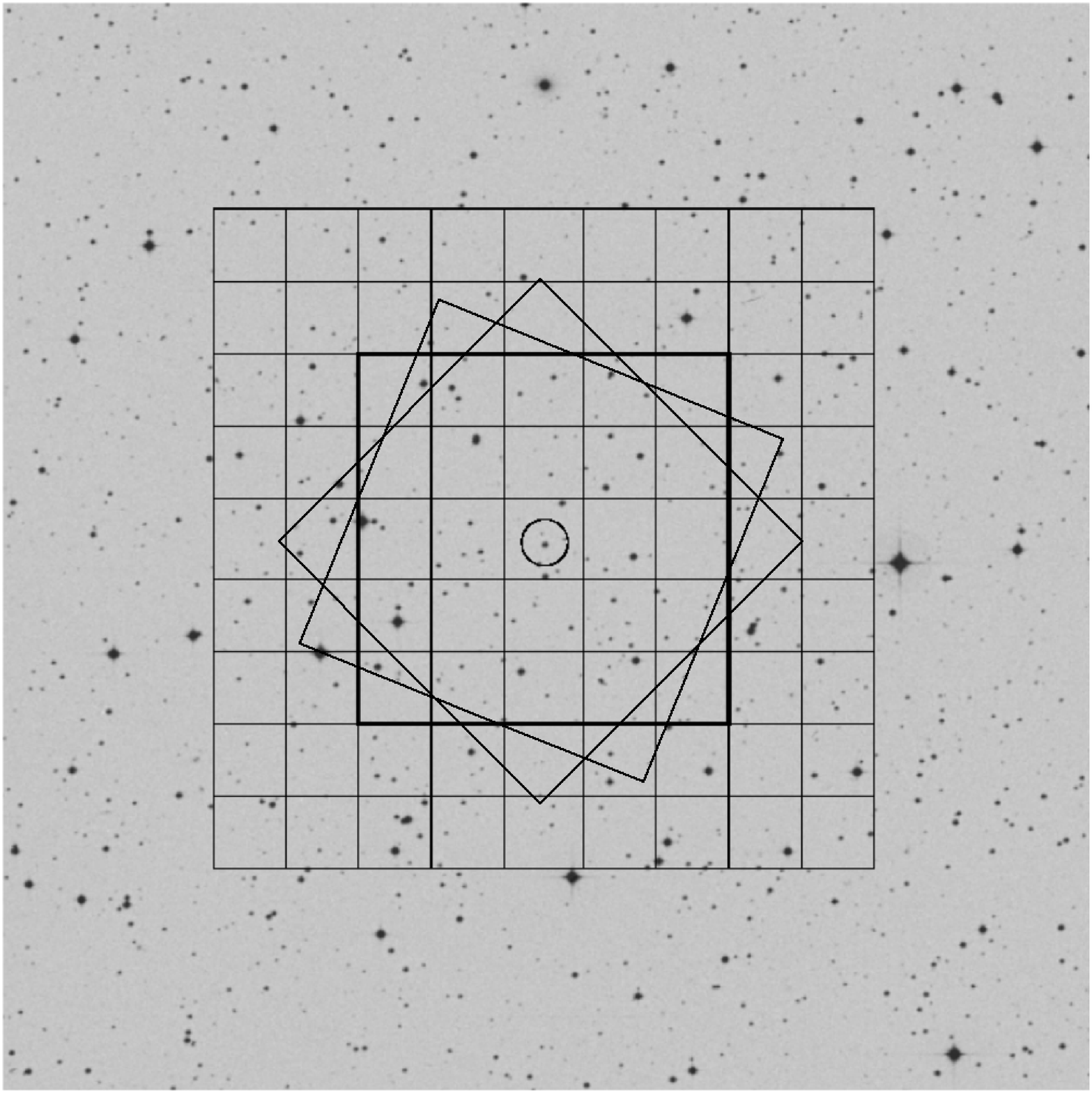}}
\caption[Pointings on the Stetson Mark~A field.]{Pointings on the Stetson
Mark~A field. The outline of the FORS1 frames are shown superimposed on the DSS
field of that region. The well observed standard star \oindex{Mark~A-S873},
which is included in all frames is marked with a circle. Note the three
different rotator angles used for the central field.  } \label{fig:grid}
\end{figure}

Our observations consisted of a $5\times5$ grid of positions on the \citeauthor{stetson}
standard field Mark~A \citep{stetson} observed at low airmass.\sxindex{Stetson
standards}  In addition, we observed one
pointing on the grid of positions with two extra position angles. The
pointings are shown in Fig.~\ref{fig:grid}.  All observations of that field
were obtained at airmasses between 1 and 1.2.

In addition, we obtained data for the three standard fields \oindex{L\,92}, \oindex{L\,113} and
\oindex{PG\,1633} observed at airmasses between 1.1 and 2.9. The FORS FOV is much smaller
than the Stetson fields. We selected subregions of the standard fields which
avoided bright stars. Unfortunately, for L113 a bright star was included in the
observed field  by mistake. This star saturated the CCD and led to bleeding,
high background and bias offsets.  For that reason, only a small fraction of
the stars on the  L113 field were useful, and 4 images  had to be completely
discarded. 

All observations were carried out in a single photometric night on July 17,
2005.

\subsection{Basic Data Reduction}

Standard subtraction of overscan region and bias frames were performed
using the IRAF ``xccdred" package.

In order to compare the quality of flatfields, we used three different
flatfields and applied them to the full set of data. They are:

\subsubsection{MASTER  FLAT:} Most reductions of FORS data use the ``master flat"
as produced by the FORS pipeline. This flat is basically the median of the flats
taken for the night of observations.  
Below we simply refer to this flat as ``\sindex{master flat}". 

\subsubsection{ILLUMINATION-CORRECTED FLAT:} As shown in the previous section,
the large scale illumination of the flats is not stable and changes from
exposure to exposure. We have therefore applied an illumination correction to
the master flat by removing its large scale variation.  We used the IRAF task
``mkillumcor" for that purpose. This task heavily smooths the master flat and
then subtracts this smooth version from the original master flat. The smoothing
kernel used by mkillumcor is a boxcar function with fixed size in the central
part of the image, and reduced size close to the edges. The minimum box size we
used was 15 pixels, and the maximum 200 pixels. A 2.5$\sigma$ clipping was used
to exclude deviant points from the computation of the smoothed image. \label{sec:illumflat}
Below we refer to this flat as ``\sindex{illumination-corrected flat}".

\subsubsection{ROTATION-CORRECTED FLAT:} Finally, as an experiment, we also used
the mean of the archive flats shown in Fig.~\ref{fig:kiloflat}.  As described
in Sec.~\ref{sec:rotator}, the input flats were selected so that flats taken
at any rotator angle are equally represented. 
We will refer to this flat as the ``\sindex{rotation angle corrected flat}".

\subsection{Measurement of Magnitudes}

Stars were identified and instrumental magnitudes were measured  using
\sindex{Sextractor}.  Based on the inspection of the growth curve, we computed aperture
magnitudes with an aperture radius of 2 arcsec, and compared them with
Sextractor's ``automag". The difference between the two magnitudes was found to
be independent of the magnitude of the stars. Because of its smaller
statistical error, the analysis was done using the ``automag".

\section{Zero Point Variation Across the FORS1 Detector}\label{sec:secondorder}

The doubt about the quality of the flatfields  discussed in Sec.~\ref{sec:flat}
and  raised by the FWII report warrants taking  a closer look at any
variations of the magnitude zero point across the detector when using the
master flat. The goal is to derive a correction to the used flatfield similar
to the one proposed by FWII to improve the accuracy of the flatfields.  In
addition, we aim to find a quantitative estimate of the accuracy of the final
adopted flatfield.

\subsection{25 Points of Light}

\begin{figure}[!h]
\centerline{\includegraphics[width=6.8cm,angle=0,clip]{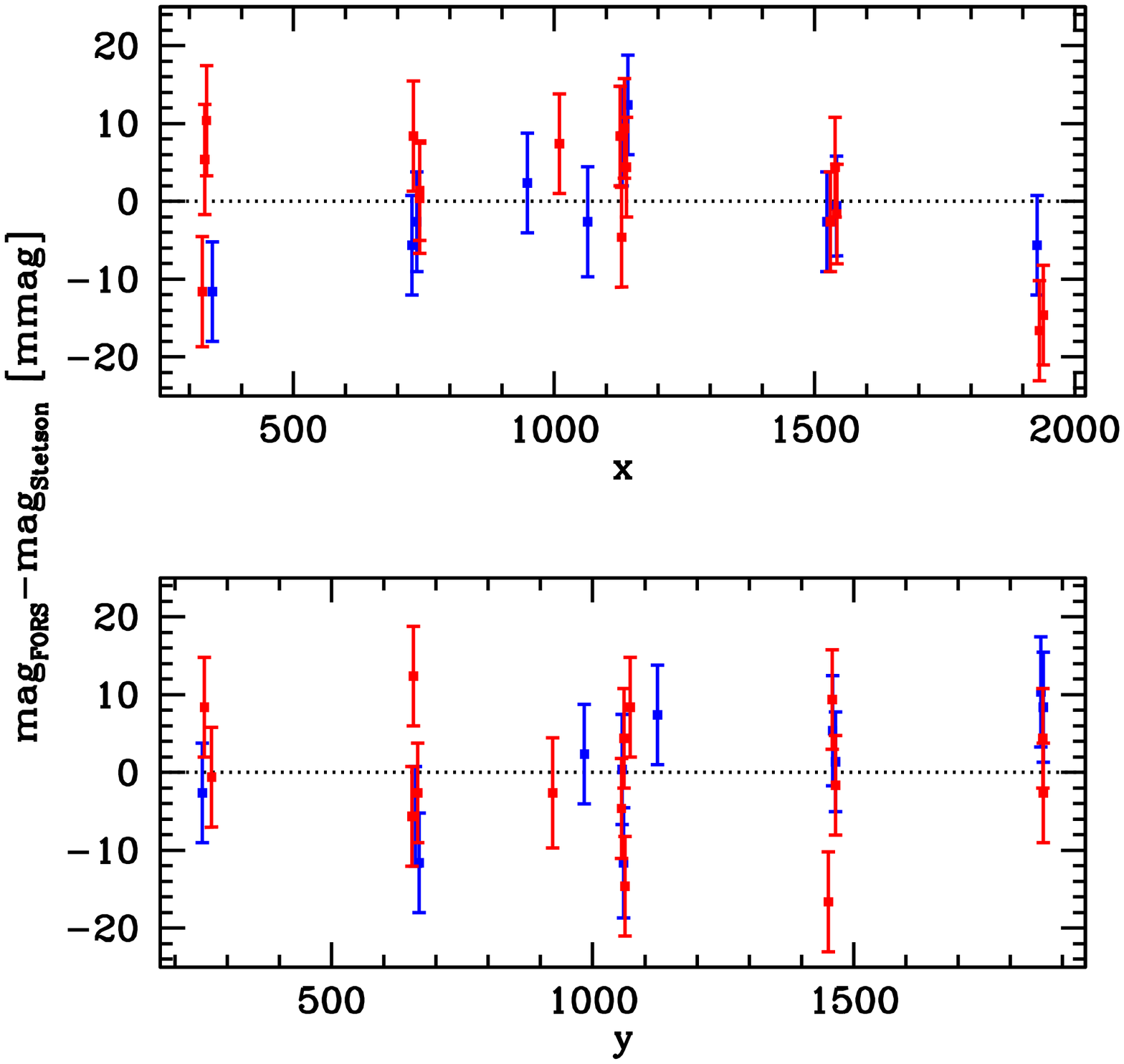}\hspace{-2mm}\includegraphics[width=6.8cm,angle=0,clip]{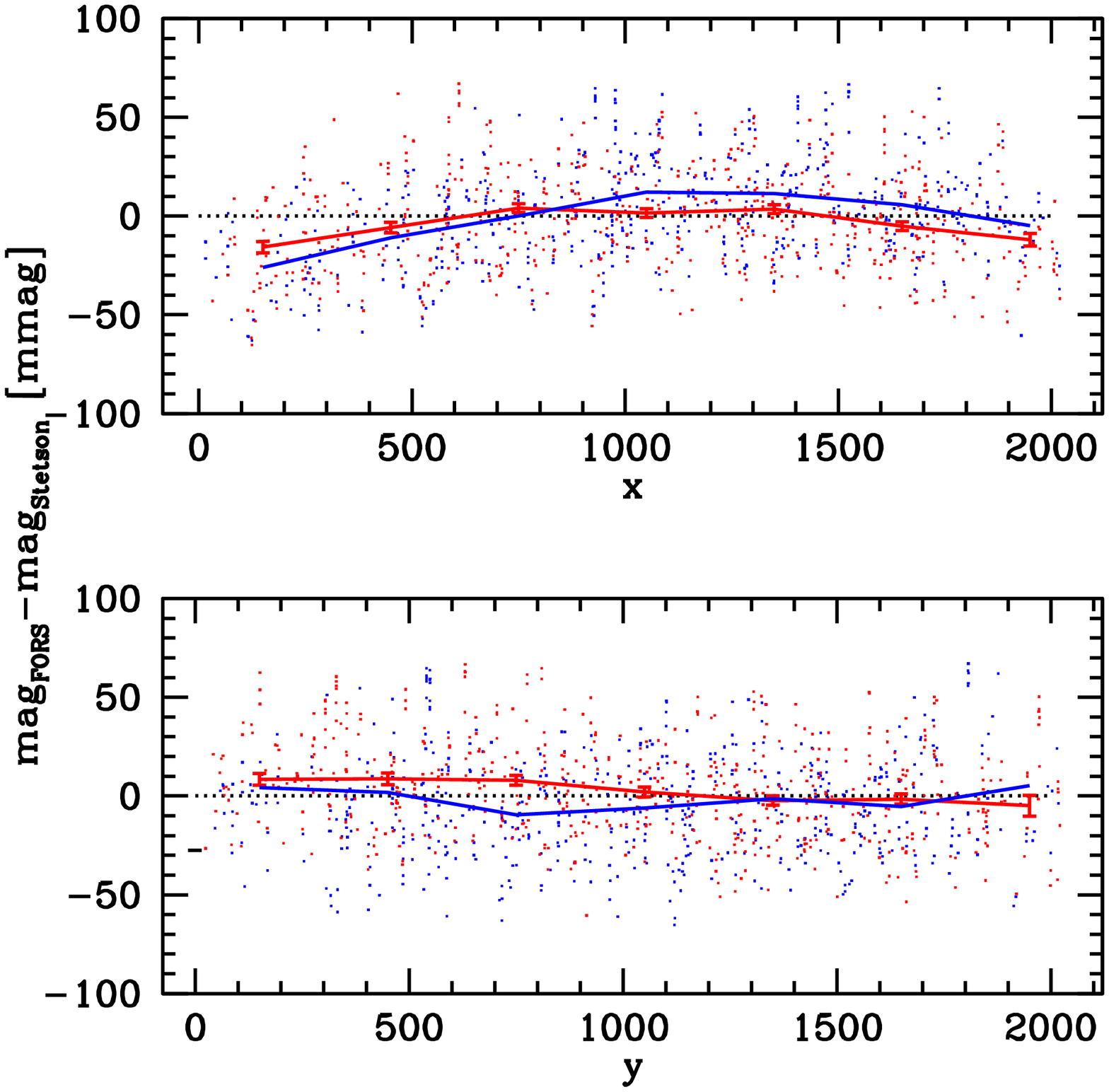}}
\vspace{-2cm}
\caption[Relative instrumental magnitude as a function of
position on the CCD.]{{\sl Left:} Data for 
Mark~A-S873.  The colour of the points in the upper panel indicates the 
$y$-coordinate of the star  in each observation. For $y>1024$,
points are red, and for $y<1024$, points are blue. Similarly, in the lower
panel,   red stands for $x>1024$, and blue for  $x<1024$. Error bars are statistical errors. 
{\sl Right:} Data for all Stetson stars in the Mark~A field are plotted. Solid lines
show the data binned in 100 pixel wide bins. 
   } \label{fig:25pol} \end{figure}

The dithered observations were planned with the specific intent of placing one of
the Stetson standard stars, namely Mark~A-S873,   on a grid of positions on the
CCD (see Fig.~\ref{fig:grid}). This approach is often nicknamed the ``1000
points of light" approach, but we call it more modestly the ``25 points of
light".  The simplest and most direct way to investigate relative \sindex{zero point
changes} with such data is to plot the relative instrumental magnitudes
of Mark~A-S873 as a function of position.  Such a plot is shown in
in the left panel of Fig.~\ref{fig:25pol}. It can  be seen that any relative photometric errors
within the part of the detector sampled by our grid  are on the order of
30$\,$mmag or less. The sensitivity achieved with this analysis is insufficient
to convincingly detect flatfield variations. 

\subsection{1000 Points of Light}\label{sec:1000pt}

Better statistics than in Fig.~\ref{fig:25pol} can be obtained by including all
stars with magnitudes listed by Stetson. About 1000 magnitudes of Stetson stars
have been measured from our data set. The right panel in Fig.~\ref{fig:25pol}
shows the difference between instrumental magnitudes and Stetson magnitudes as
a function of position on the detector. No bandpass correction was applied. The
scatter of individual points in this plot is larger than the scatter in
Fig.~\ref{fig:25pol} because of errors in the Stetson magnitudes and because
differences between the FORS1 and Stetson's effective filter shapes makes the
zero point of stars depend on colour.  This is more than compensated by the
larger number of measurements. This larger data set clearly detects some
deficiencies in the flatfield  with  total peak-to-peak error in relative
photometry, within the inner part of the detectors, of about 30$\,$mmag.

\subsection{Many Points of Light}
\subsubsection{\sindex{Flatfield Correction Factor} $f(x,y)$}

The images of our data set contain many more stars suitable for photometry than
the ones listed by Stetson.  The large number of stars allows to simultaneously
fit for relative zero points of each image, the relative magnitude of  each
star and zero point variations across the field.  FWII describes a method to
find such a solution from a set of measured magnitudes on a set of dithered
images.  The power of this approach comes from the much larger number of stars
which can be used to improve the statistics compared to the previous
approaches. By contrast, the number of free parameters (one for each star and
observed field) increases only modestly. 

\begin{figure}[!h] 
\centerline{\includegraphics[width=5cm,angle=0,clip]{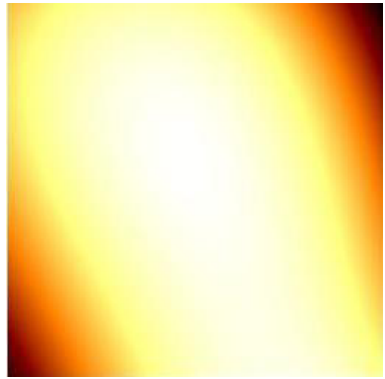}}
\caption{$R$-band flatfield correction frame. }
\label{fig:flat1} \end{figure}

FWII defined a flatfield correction factor $f(x,y)$, so that each measured
magnitude of a star  on any of the  images can be written as

\begin{equation}
m_{\nu\mu} = M_\nu + z_\mu - f(x,y)
\end{equation}

\noindent where $M_\nu$ is the magnitude of star $\nu$ within the chosen band,
$m_{\nu\mu}$ is its instrumental magnitude measured on image $\mu$, and $z_\mu$
is the zero point of image $\mu$.  Following FWII's approach, we used a
polynomial to model $f(x,y)$, 

\begin{equation}
   f(x,y)= \sum_{i=0}^o\sum_{j=0}^{o-i} p_{ij}x^iy^j.
\end{equation}

\noindent The formalism to compute $f(x,y)$ is described in Appendix~\ref{sec:form}

To estimate the uncertainties  in the correction frame, we have carried out
Monte-Carlo simulations in the following manner. First, we added normally
distributed random errors to the measured magnitudes of each star.  The
standard deviation  of the Gaussian was chosen to be identical to the error
estimate in the actual measured magnitude. We created a total of 100
artificial data sets in this manner, and fitted  $f(x,y)$ for each of them. We
then computed the rms from all artificial data sets for each pixel. 

\subsubsection{Results}\label{sec:corr}

The resulting $f(x,y)$ flatfield correction is shown in Fig.~\ref{fig:flat1}.
The peak-to-peak flat-fielding error at the position of the observed stars is
about 30$\,$mmag. The peak-to-peak flatfield correction over the whole field is
about 50$\,$mmag. However, over a large fraction of the detector, the
corrections are smaller than 10$\,$mmag and the rms over the whole frame
excluding a strip 200 pixels wide along the edge  is only about 4$\,$mmag.
Therefore, while flat-fielding problems on FORS1 might result in errors larger
than our stated goal of 3\% photometry for individual stars,
statistically for random positions on the detector, the errors are much
smaller.  A different strategy for achieving accurate relative photometry with
FORS1 is to concentrate on the centre part of the detector. For example, within
the central $4\times4\,$arcmin of the detector, roughly one third of the
detector area, the difference between the minimum and the  maximum of the
correction factor is about 13$\,$mmag, and the rms is 2$\,$mmag.

\subsubsection{Comparison with FWII}

\begin{figure}[!th] 
\centerline{\includegraphics[width=13cm,angle=0,clip]{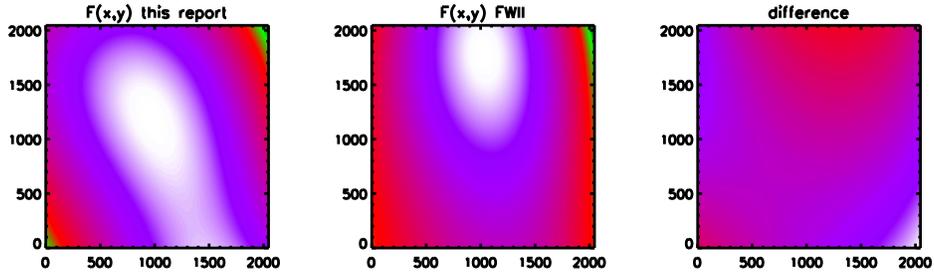}}
\caption[Comparison between flatfield correction frames.]{Comparison between the flatfield correction frames derived in the current work
and FWII. {\sl Left}: $f(x,y)$ from {\em FAP}, {\sl center}:
 from FWII, {\sl right}: the difference. The colour scale in
all three panels is identical. } \label{fig:compare_moeller} \end{figure}

\begin{figure}[!h] 
\centerline{\includegraphics[width=8cm,angle=0,clip]{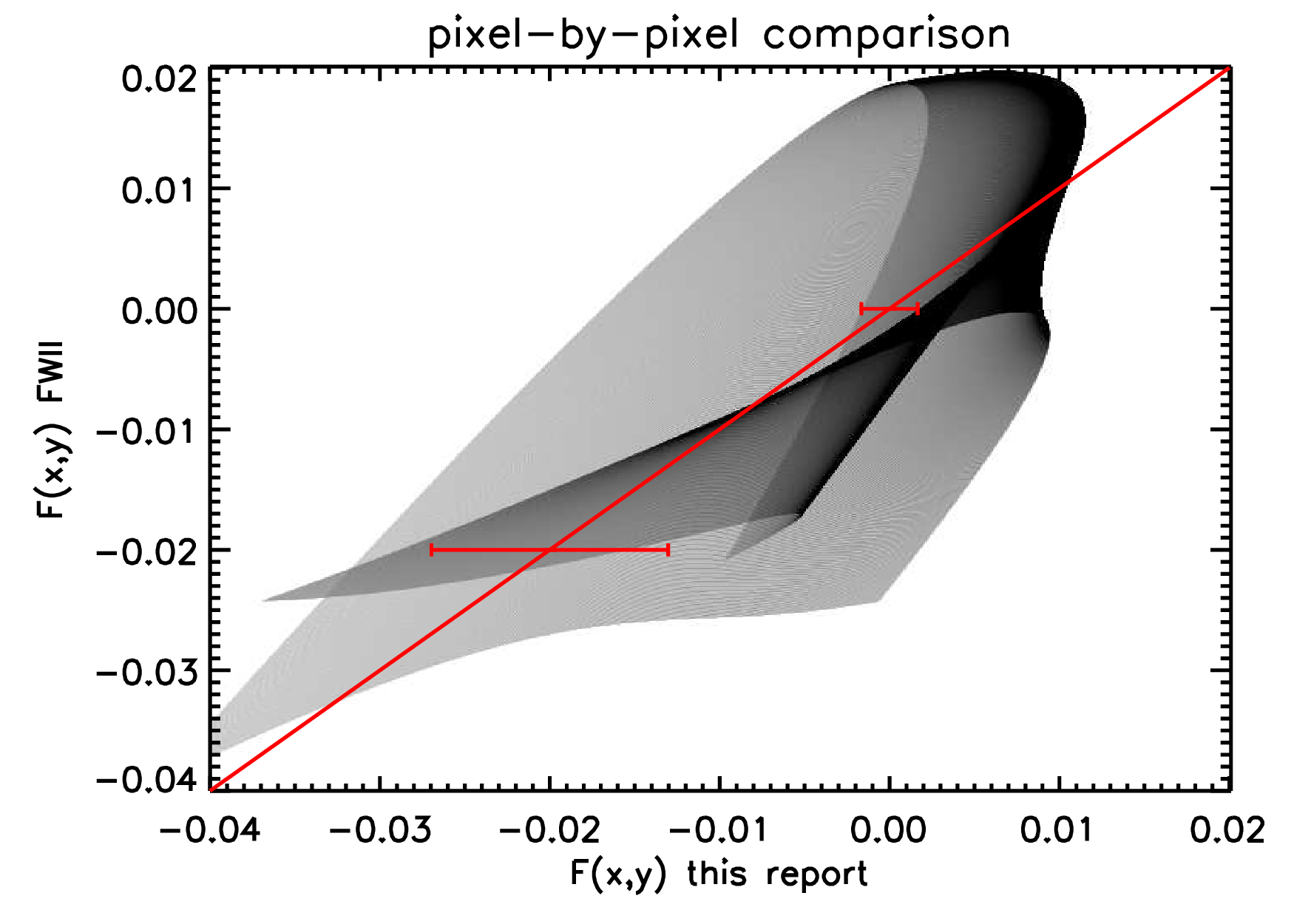}}
\caption[Pixel-by-pixel comparison of $f(x,y)$.]{Pixel-by-pixel comparison of
$f(x,y)$ found by FWII and in the current project. The gray level indicates the
number of pixels with the corresponding combination of values found by the two
fits. The red line illustrates a one-to-one correspondence of the plotted
values.  The error bars are the minimum and maximum rms uncertainty for
$f(x,y)$ estimated from the \sindex{Monte-Carlo simulations}. The differences between 
the two frames are comparable to the uncertainty in the fits.  }
\label{fig:compare_moeller_pixel} \end{figure}

FWII used a similar procedure to the one used for the current
work.  In Fig.~\ref{fig:compare_moeller} and~\ref{fig:compare_moeller_pixel},
we compare the results of our fit to the one in FWII. It can be seen that there
is a good correlation between the two flatfield correction frames from data
taken more about 15 month apart. The differences  between the two determinations
of $f(x,y)$ are similar in magnitude to the error estimates in $f(x,y)$.
This suggests that there is a long term
 stable flatfield correction which can be applied to improve the photometric
quality of images taken with FORS1 in the $R$-band. 

\section{Improving the Master Flat}

\begin{figure}[!h]
\centerline{\includegraphics[width=6.7cm,angle=0,clip]{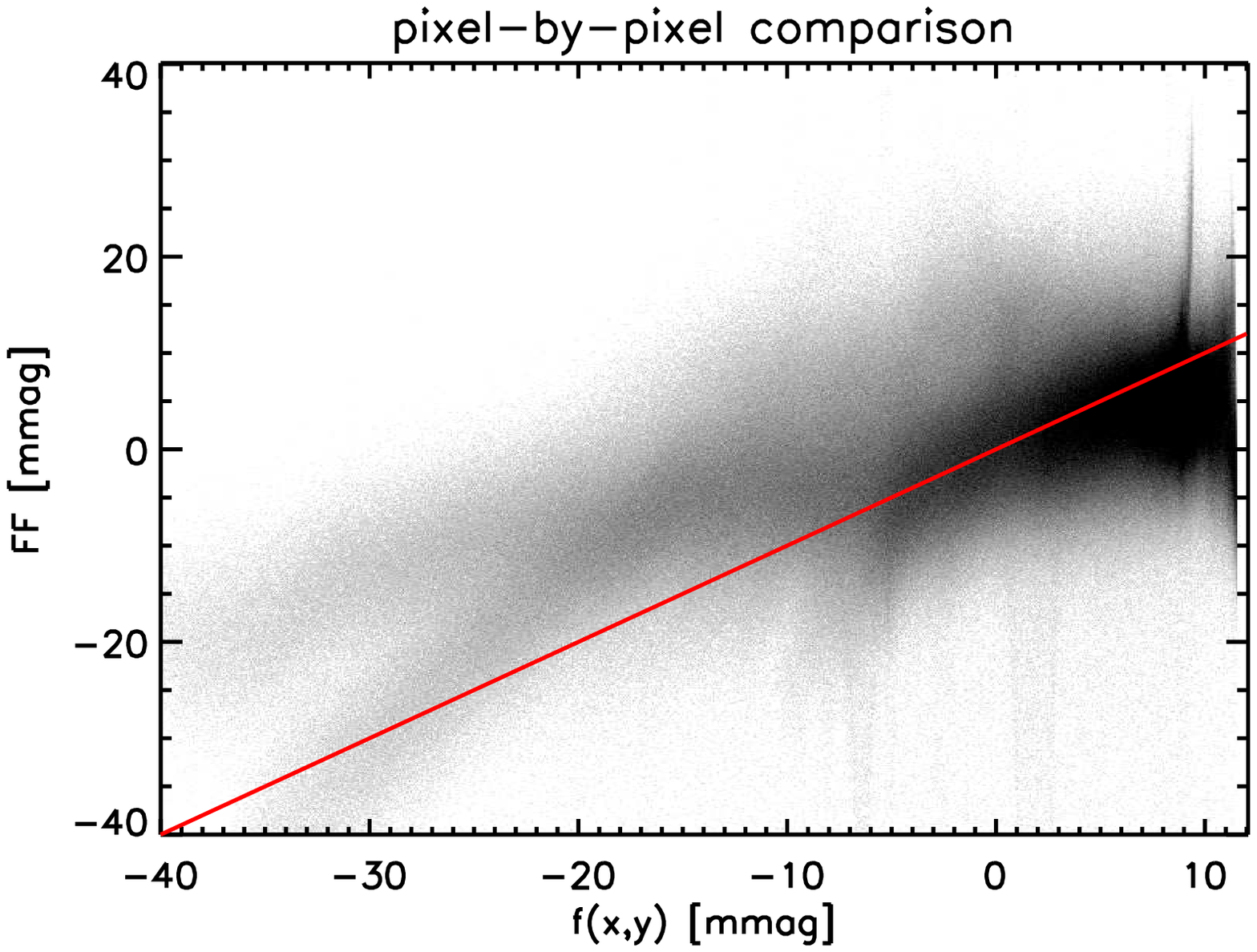}\includegraphics[width=6.7cm,angle=0,clip]{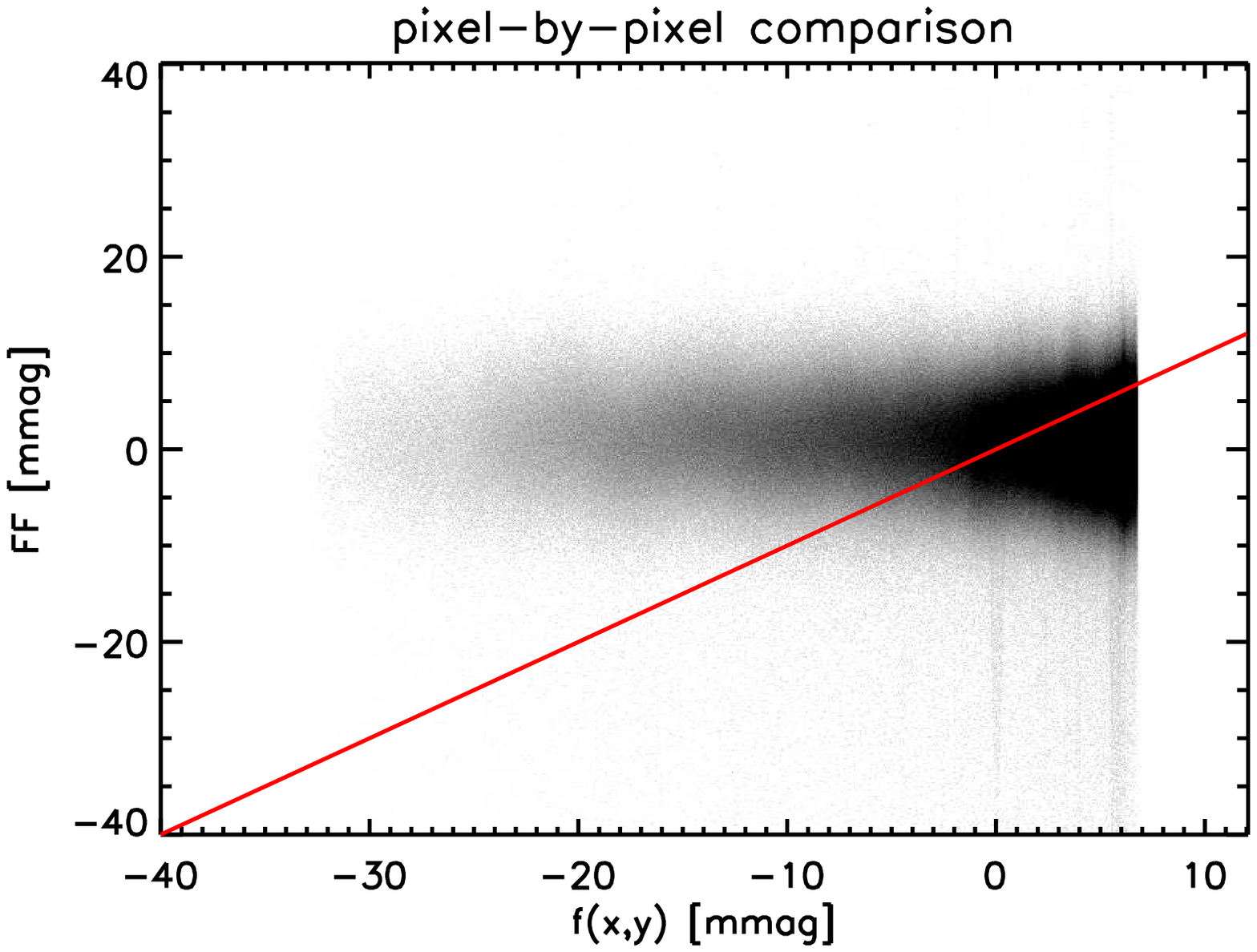}}
\caption[Pixel-by-pixel comparison of $f(x,y)$ with the flats]{Pixel-by-pixel
comparisons of $f(x,y)$ with the flatfield (FF) used to process the images
before fitting $f(x,y)$. {\sl Left}: master flat versus derived $f(x,y)$, {\sl
right}: illumination-corrected flat versus $f(x,y)$.  } \label{fig:compare_FF}
\end{figure}

In Sec.~\ref{sec:rotator} we have shown that a substantial component of the
structure in the master flatfield rotates with the adaptor rotator. It is
unlikely that any feature in the sensitivity map, i.e. the ``true" flatfield,
rotates.  Therefore, it is most likely that the rotating feature is a defect in
the master flats, e.g. caused  by light scattered on some structure connected
with the rotator. In this case, the derived flatfield correction should
compensate for some of the structure found in the master flat. In the left
panel of Fig.~\ref{fig:compare_FF}, we compare the master flat with the derived
$f(x,y)$ on a pixel-to-pixel basis.  We find that there is a significant
correlation between the two frames. This suggests that the master flat could be
improved simply by removing its large scale pattern.

This motivated us to create the illumination-corrected flat described in
Sec.~\ref{sec:illumflat}  We  used the images flatfielded with this modified
flat to  re-measure magnitudes and re-derive the flatfield correction factor. A
pixel-by-pixel comparison of the illumination-corrected master flat with the
re-derived correction factor is shown in the right panel of
Fig.~\ref{fig:compare_FF}.  It can be seen that any correlation between flat
and correction factor has successfully been removed, and that the peak-to-peak
values of the correction factor have become smaller.  This demonstrates that
the master flat can be improved by this simple procedure. We also tested the
same procedure on the $I$-band data, and on the FWII data and found similar
results.

\begin{figure}[!h] 
\centerline{\includegraphics[width=6.7cm,angle=0,clip]{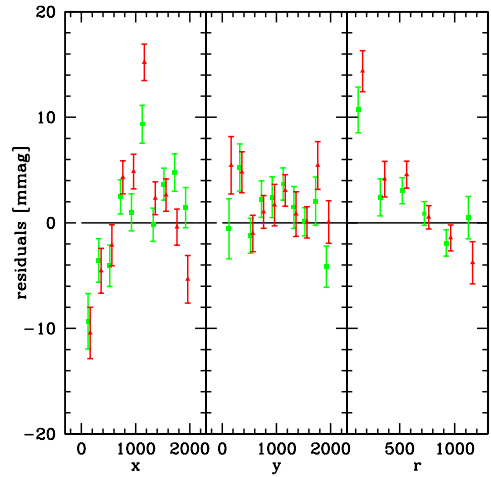}}

\caption[Residuals from fit of photometric solution as a function of detector
position.]{Residuals from fit of photometric solution as a function of detector
position. Left, centre and right panels show the mean residuals as a
function of column $x$, row $y$ and distance from the detector centre $r$.  The red
points are derived   using the regular master flat, and the green points used
the illumination-corrected master flat. No flatfield correction $f(x,y)$ has
been applied. Error bars are the errors of the means based on error estimates
for  the measured magnitudes and listed errors of the standard stars.  Note
that the mean of all the  residuals is by construction zero. In the leftmost
panel, the inner point contains fewer stars because it is based on a small area
on the detector.  } \label{fig:residuals} 
\end{figure}

The standard stars on the images can be used to verify that the flat-fielding
is indeed improved by this procedure. For that purpose, we derived photometric
solutions from the standard star measurements  as described in
Sec.~\ref{sec:abs}, but without using the flatfield correction factor $f(x,y)$.
The residuals as a function of detector position for the case of the regular
master flat and that of the illumination-corrected flat are compared  in
Fig.~\ref{fig:residuals}. It can be seen that the illumination correction
improves flatfield errors by as much as 50\% in the centre of the field.
However, it also shows that even using the illumination-corrected flats,
significant flat-fielding errors remain and the flatfield correction procedure
is still needed to reduce residual flatfield errors to values below 1\%.

We have  also tried the same procedure using the rotation-corrected flat shown in
Fig.~\ref{fig:kiloflat} but found no improvement over the standard master flat.
We therefore will not use that flat in the further analysis.

\section{Absolute Photometry}\label{sec:abs}

\subsection{Photometric Quality of Night}\label{sec:night}

A crucial requirement for {\em FAP} was  that observations were carried out
under photometric conditions. The judgement whether a night on Paranal is
photometric is done by the weather officer. This judgement is based on zero
points derived from observations with the various imaging cameras,  and
inspection of the sky.  Inspection of the sky is carried out by eye and with
the help of MASCOT, a sky monitor which delivers optical images of the whole
sky every three minutes.  For {\em FAP}, the photometric quality of the night
can be judged from the collected data.  This might not be the case for science
programmes which take significantly  fewer calibration observations during a
night.  If such a programme requires  photometric conditions, it is essential
that the observer can judge the quality of the night objectively.  This is of
particular importance for service mode observations.

\begin{figure}[!t] 
\centerline{\includegraphics[width=6.7cm,angle=0,clip]{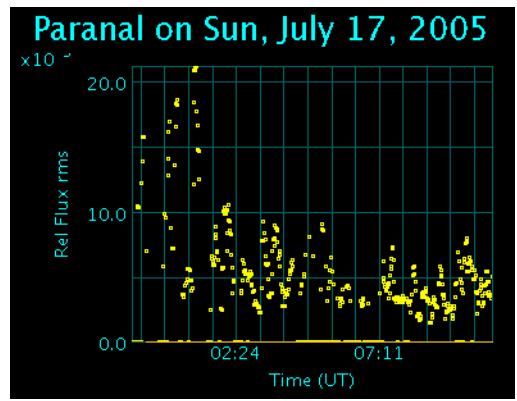}}

\caption{The rms flux as measured by the ASM monitor. 
}
\label{fig:nightquality} \end{figure}

\begin{figure}[!t] 
\centerline{\includegraphics[width=6.7cm,angle=0,clip]{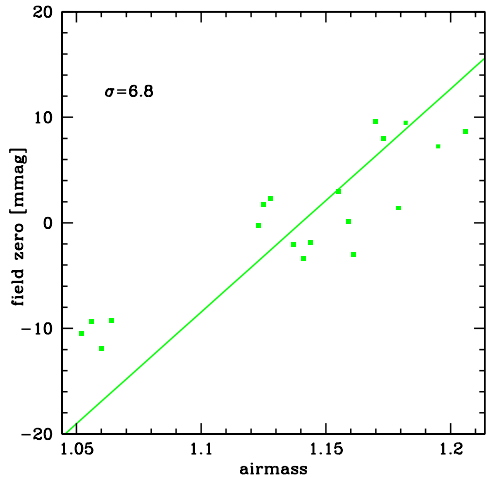}}
\caption[Relative  zero points of individual exposures of the Mark~A field as a
function of airmass.]{Relative  zero points of individual exposures of the
Mark~A field as a function of airmass as determined from the simultaneous fit
of $f(x,y)$, relative magnitudes of stars and the exposure zero points.  No
standard magnitudes were used in deriving these data points.  The solid line is
the slope of the extinction correction as determined from the fit to Stetson
magnitudes of stars. The rms of the  deviations from this line is 6.8$\,$mmag.
} \label{fig:zeroair} \end{figure}

One of the tools to judge the quality of the night  is the ``VLT Astronomical
Site Monitor" (ASM\footnote{\tt http://archive.eso.org/asm/ambient-server}).
Figure \ref{fig:nightquality} shows the ASM flux fluctuations during the
course of the night when the {\em FAP} observations were carried out. All {\em
FAP} images were taken after UT 2:20 when the rms of the fluctuations was less
that 10$\,$mmag, and the mean rms fluctuation during that time was about
7$\,$mmag.  This rms from the ASM monitor can be compared to the flux
fluctuations derived from the observations.

The procedure described in Sec.~\ref{sec:1000pt} yields the relative zero point
for each image of the Mark~A field. These zero points are already corrected for
flat-fielding errors and each one of them is based on the weighted average of
more than 1000 stars.  The random errors of the relative zero points are
therefore extremely small, about one mmag.  Changes in the relative zero points
are therefore a highly accurate measure of changes in the extinction between
different images. 

Figure~\ref{fig:zeroair} shows the relative magnitude zero points for the
Mark~A images as a function of airmass. The error estimate for each of the
points based on the measurement errors is smaller than the point size. Also shown
is the slope of the extinction curve derived from fields taken at higher
airmasses, the details of this determination will be discussed in
Sec.~\ref{sec:ext}  It can be seen that the slope of the extinction curve
is in excellent agreement with the variations of the zero point as a function
of airmass. The rms scatter of the zero points around the extinction curve is
6.8$\,$mmag. This experiment confirms the excellent photometric quality of the
night completely independent of any standard star magnitudes.

The scatter of 6.8$\,$mmag is a good measure of the fluctuations in the
extinction within the 10$\,$sec exposures.  Its  value is similar to the flux
rms measured by the  ASM monitor.  It is tempting to conclude that the rms from
the ASM can be used as a proxy for expected rms fluctuations of the zero point.
Whether this is indeed the case warrants further investigation. For example,
one area of concern is its sensitivity to seeing changes.  

\subsection{Photometric Solution }\label{sec:phot}

The method used to find the \sindex{flat-fielding correction} can easily be modified
to find a photometric solution from the current data set. Instead of using an
arbitrary zero point for each star and each field, the photometric zero point, 
colour terms and extinction coefficients are fitted. Specifically, we assumed 
that the instrumental magnitudes $r$ and the Stetson magnitudes $R$ and $I$ are
related as

\begin{equation}
R-r= z + e\cdot x + a\cdot (R-I) + c\cdot (R-I)\cdot x   \label{equ:ext}
\end{equation} 

\noindent where $x$ is the airmass and $z$, $e$, $a$ and $c$ are parameters determined by
the fitting.  Those parameters were fitted simultaneously with the flat-fielding correction. 
The formalism is described in Appendix~\ref{sec:form_mag}.

We compared this solution to a separate fit of the correction frame $f(x,y)$
followed by a fit of   the photometric solution.  We found no differences in
the results.  All results in this section are based on  the
illumination-corrected flatfields and the additional application of the flat
field correction derived from all detected stars. 

Colour coefficients were fit using 
\begin{equation}
r-i= c_z + c_e\cdot x + c_a\cdot (R-I) + c_c\cdot (R-I)\cdot x + c_d\cdot(R-I)^2,\label{equ:color} 
\end{equation}

where $c_z$, $c_e$, $c_a$, $c_c$ and $c_d$ are the parameters of the fit.

\begin{figure}[h]
\centerline{\includegraphics[width=6.7cm,angle=0,clip]{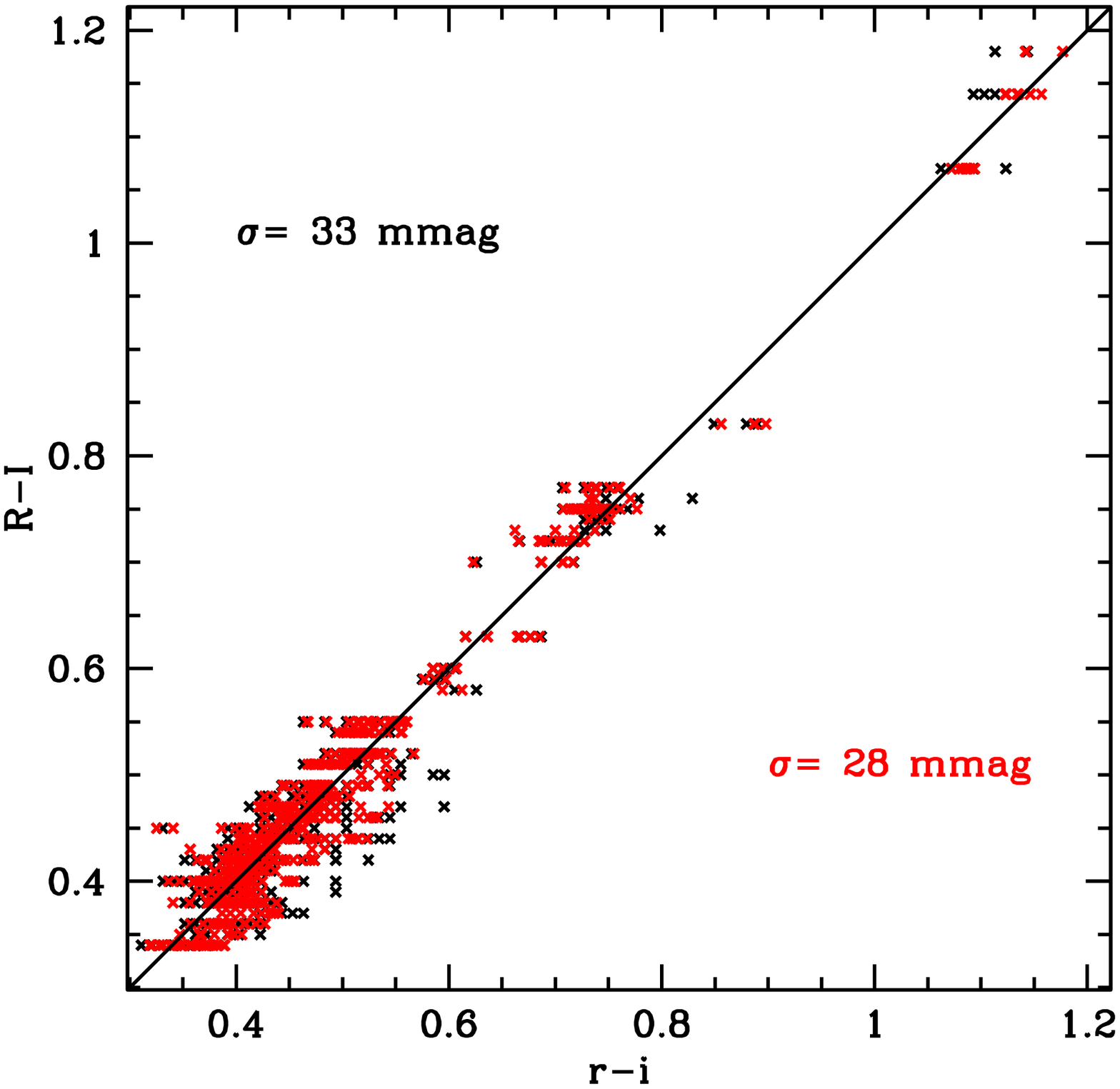}
            \vspace{1mm}
            \includegraphics[width=6.7cm,angle=0,clip]{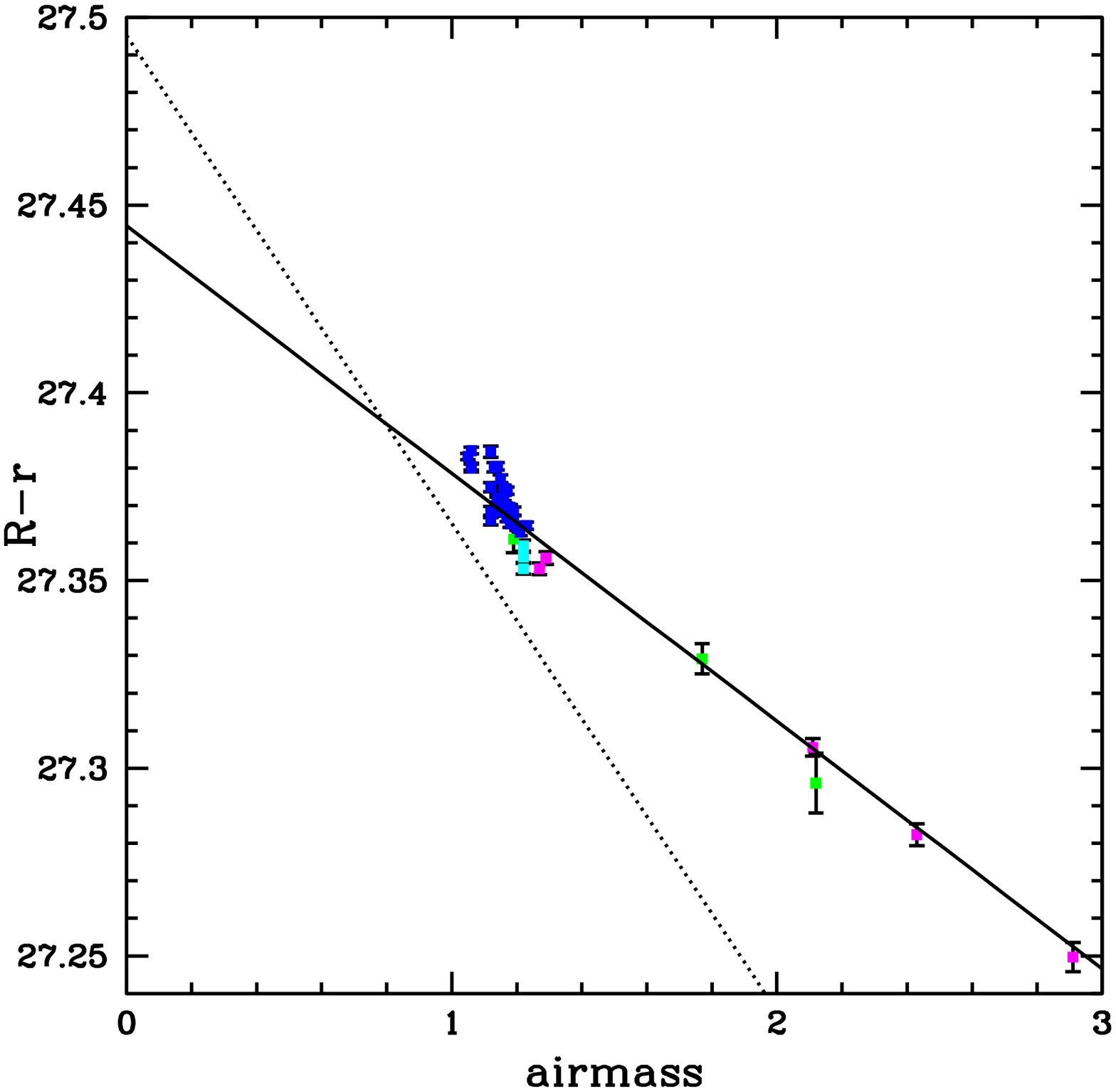}}
 \caption[Extinction and Colour
solutions.]{Extinction and Colour
solutions. {\sl Left:} 
{$R-I$ colours listed by Stetson versus the  colours
computed from the {\em FAP} data. Points in black use the colours based only on
a fit of the zero point and slope of the colour correction, whereas red points
use the full airmass and quadratic terms.  }
{\sl Right:} 
 $R-r$ corrected for colour and colour-dependent extinction as a function of
airmass. Each point is the weighted mean of all stars in one image. The colour
of each point indicates the Stetson field from which the point was derived. The
codes are: blue: Mark~A, green: L\,113, magenta: PG\,1633 , cyan: L\,92. The solid
line is the fit to the data points and the extrapolation to zero airmass is
shown to illustrate the magnitude zero point.  For comparison,   the
photometric zero point  and assumed extinction from the QC pipeline for that
night is shown as a dotted line.  } \label{fig:ext} \end{figure}

\subsection{Results}\label{sec:ext}

\subsubsection{Colour Solution}

The linear component of the colour solution (Eq.~\ref{equ:color}) is shown in
the left panel of Fig.~\ref{fig:ext}.  The scatter in the fit is about
28$\,$mmag. This fit can be used to replace $R-I$ in   Eq.~(\ref{equ:ext}) with
its measured values.   The uncertainty in the true colour adds less than
3$\,$mmag to the random  error of the final $R$ magnitude.

\subsubsection{Extinction Solution}

The resulting extinction solution is shown in right panel of
Fig.~\ref{fig:ext}.  The ESO Quality Control (QC) group derives a photometric
zero point assuming an extinction for each night.  This QC zero point and
extinction for that night are also shown in Fig.~\ref{fig:ext}. There is a
small offset between the normalisation of the QC extinction curve and the one
derived here at airmass around 1.2.  This offset  might be due to slight
differences in the normalisation of the flatfields, differences in the
apertures used to measure magnitudes, and/or differences in the colour
coefficients.  However, the assumed extinction in the QC procedure introduces
an additional error in the zero point which is much larger than the differences
at airmass around~1.2.  The extinction varies substantially from night to
night, even when the nights are photometric.  Therefore, zero points derived
using a mean extinction depend on the airmass of the measured standard field
and are not useful for photometry. The true photometric zero point above the
atmosphere as derived from extrapolation of the extinction curves probably
varied much more slowly than the night-to-night variations of the extinction.
For this reason, when only a single photometric standard observation is
available in a given night, the best practise is to derive the extinction
coefficient for that night by assuming the zero point has not changed from the
previous determination \citep[cf.  e.g.  ][]{harris}.

\begin{figure}[!h]
\centerline{\includegraphics[width=6.7cm,angle=0,clip]{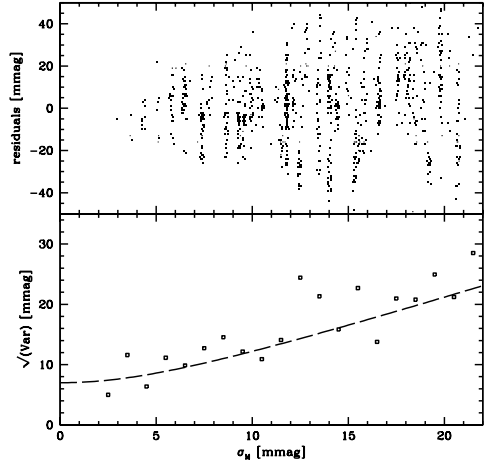}} 
\caption[Residuals as function of the estimated magnitude error.]{Upper panels:
Residuals as function of the estimated magnitude error.  Lower panel: The
$\sqrt{{\rm Var}}$ of the residuals as a function of magnitude error. The
superimposed line corresponds to  Var$= (7 {\rm mmag})^2+ \sigma_M^2$.  
\label{fig:residuals_flat}} \end{figure}

\subsubsection{Residuals and Error Budget}\label{sec:err}

Sextractor computes error estimates $\sigma_s$ for each measured magnitude. The
error includes the contribution of the  readout noise and Poisson noise, both
for the pixels used to compute the stellar flux and for those used to estimate
the local background. The error estimates ranged from 2 to 30$\,$mmag.
\citet{stetson} and \citet{stetson_fields} list error estimates $\sigma_{st}$
for individual standard stars based on repeated observations in different
nights. The error estimates for the stars used in this analysis  range from 2
to 20$\,$mmag. By comparing these error estimates to the residuals of the
extinction solution, we can find an external estimate of  the combined effect
of {\em all sources of  errors} not included in $\sigma_s$ and $\sigma_{st}$.
For this purpose, we plotted the residuals from the extinction solution as a
function of the error estimate $\sigma_M$ for each $R-r$. The plot is shown  in
the upper panel of Fig.~\ref{fig:residuals_flat}. The error estimate $\sigma_M$
was computed as $\sigma_M= \sqrt{ \sigma_s^2 + \sigma_{st}^2 }$.  It can be
seen that the scatter in the residuals for small estimated errors is less than
10$\,$mmag and increases for larger $\sigma_M$.  The lower panel of
Fig.~\ref{fig:residuals_flat} shows the rms of the residuals binned by error
estimates. A source of scatter in addition to the error estimate has to be
assumed to account for the scatter residuals. Assuming  this additional scatter
$\sigma_{a}$ is independent of magnitude, the total scatter in the
residuals $\sqrt{\rm VAR}$  can be modelled as 

\begin{equation} {\rm VAR}^2 = { \sigma_s^2 + \sigma_{st}^2 +\sigma_a^2 }
\end{equation}

\noindent The lower  panel of  Fig.~\ref{fig:residuals_flat} shows that a
value of $\sigma_a\approx7\,$mmag is consistent with the residuals.

Sources for $\sigma_a$ include extinction fluctuations $\sigma_e$, colour transformation errors $\sigma_c$ and residual flat-fielding errors $\sigma_{ff}$. 
The total error estimate $\sigma_t$ for our magnitude measurements becomes
therefore
\begin{equation}
 \sigma_t  = \sqrt{ \sigma_s^2 +  \sigma_a^2 } \\
           = \sqrt{ \sigma_s^2 +  \sigma_e^2 +\sigma_c^2+\sigma_{ff}^2} \\
\label{equ:errs}
\end{equation}

In Sec.~\ref{sec:night} we found that  $\sigma_e\approx7\,$mmag, and in
Sec.~\ref{sec:phot} we estimated that $\sigma_c\approx3\,$mmag. Using 8$\,$mmag
as the upper limit for $\sigma_a$, we find from Eq.~(\ref{equ:errs}) an upper
limit on residual flat-fielding and other sources of errors of about $3\,$mmag.
We therefore conclude that extinction variations, statistical errors and errors
in the standard magnitudes   account for most of the residuals of our
photometric solution.

\subsection{How many Standard Field Observations are Necessary?}\label{sec:howmany}

An important goal of {\em FAP} is to find a set of guidelines on how to achieve
a photometric accuracy of 3\% or less. The photometric zero point is obviously
an important factor which determines the final accuracy of the magnitudes.  The
{\em FAP} observations contain a large number of standard stars on each
individual image, and the number of calibration images is much bigger than the
number realistically taken for the calibration of normal science observations.
An important part of the photometric  guidelines are the necessary minimum
number of  standard fields needed to achieve a certain photometric accuracy.

\begin{figure}[]
\centerline{\includegraphics[width=6.7cm,angle=0,clip]{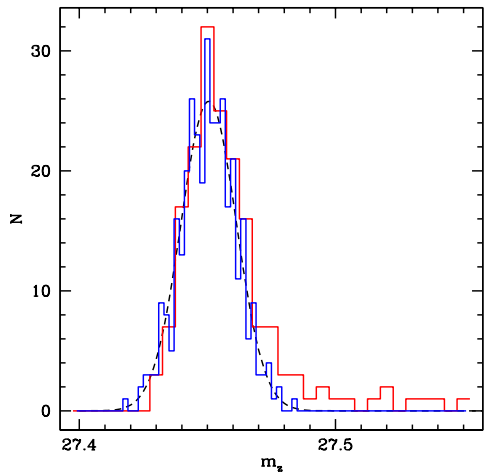}}
\caption[Residuals as function of the estimated magnitude error.]{Distributions
of zero points determined from two (red histogram) and three (blue histogram)
standard observations.  The dashed  line is a Gaussian with a $\sigma$ of
11$\,$mmag.  } \label{fig:zero} \end{figure}

{\em FAP} imaged four different Stetson fields. The magnitude and colour range,
and the consistency of  derived zero points  seems to be similar for all fields
(see e.g. Fig.~\ref{fig:ext}). In addition, we searched for and did not find
any evidence for different behaviour of the residuals as a function of
position, magnitude or colour. Therefore, there is  no evidence than anyone of the
three fields Mark~A, L92 or PG1633  is better suited for photometry than
any other. As discussed in Sec~\ref{sec:obs}, the particular region we used within
the L113 field was not optimally chosen. Excluding the L133 field from the
analysis  in this section did not change any of the conclusions. For that
reason,   we do not distinguish between the different fields in the following
discussion. 

If a night is known to be photometric, e.g. by consulting the ASM, a minimum of
2 calibration fields at different airmasses are needed to find the extinction
coefficient. The optimum strategy is that one of them is at as low an airmass
as possible, while the other one is at the highest possible airmass. A
realistic goal is to observe the low airmass field at an airmass less than 1.3,
and the high airmass field at airmass above~2. 

To estimate the errors in the zero points from sets of only two standard field
observations,  we re-computed the zero points from subsets of the {\em FAP}
data.  We used  every combination of two standard  fields  which satisfy
the above  constraints  on the airmasses.  The distribution of the resulting
zero points  is shown in Fig.~\ref{fig:zero}. The distribution has an almost
Gaussian peak but also a long non-Gaussian tail.  In about 10\% of all
cases, the errors on the resulting zero points is larger than 3\%. We
therefore conclude that observation of only two standard fields is
insufficient to photometrically calibrate a night to sufficient accuracy.

We then repeated the experiment using 3 standard fields. In each case, only one
of the three fields was chosen to be at airmass lower than 1.3, because the
gain from additional low airmass fields was judged to be small.  The resulting
distribution of zero points  is plotted in Fig.~\ref{fig:zero}. Also
shown is a Gaussian with the same mean, standard deviation and normalisation as
the zero point distribution.  It can be seen that the distribution resembles
closely the Gaussian with a standard deviation of 11$\,$mmag. In contrast to
the previous experiment with only two standard fields, all zero point errors
are less than 3\%. This result strongly suggests that 3 photometric
standard fields, chosen with the strategy outlined above, lead to an accuracy  of
about  11$\,$mmag.

The $\chi^2$ per degree of freedom of the deviation between the Gaussian fit
and the histogram in Fig.~\ref{fig:zero} based on count statistics is about
0.85. This means than the distribution of zero points when using 3 different
standard fields very closely follows a Gaussian distribution. The error budget
discussed in  Sec.~\ref{sec:err} implies that the dominant error on the mean
magnitude of all stars in any of the standard fields  are fluctuations in the
extinction which affects all stars of an image in the same way. The only way to
improve the magnitude zero point is therefore to increase the number of
independent exposures.   The fact that the distribution of the residuals shown
in Fig.~\ref{fig:zero} is normal suggests that adding more stars will improve
the accuracy of the zero point, and the final error in the zero point
$\sigma_Z$ is 

\begin{equation}
\sigma_Z \approx 11\,{\rm mmag} \cdot \sqrt{\frac{3}{n_f}}\label{equ:zeroerr}
\end{equation}
where $n_f$ is the number of standard field observations. This formula should apply if the number of standard stars in each field is large enough so that 

\begin{equation}
  \sqrt{\frac{1 }{\sum{ \frac{1}{\sigma_{\rm STD}^2} }}} \ll 11\,{\rm mmag}\label{equ:numstd}
\end{equation}
and the exposures sample the airmass between 1 and 2 uniformly.  For a typical
magnitude uncertainty $\sigma_{\rm STD}$ of 10$\,$mmag, about 30 or more
standard stars per field are needed to satisfy Eq.~(\ref{equ:numstd}). This is
one of the reasons to use the Stetson standard fields as opposed to fields with
fewer known magnitudes.   Unfortunately, the {\em FAP} data do not include a
sufficient number of independent observation to test
formula~(\ref{equ:zeroerr}) for $n_f$ larger than 3.

The zero point error is a systematic additive error which  affects all derived
magnitudes in the same manner. The exact impact of such an error depends on the
science application.  In most cases, a programme with a stated goal to achieve 3
percent photometry requires that the systematic error is significantly less than
3\%.  The  10$\,$mmag accuracy for the zero point might therefore not be
sufficient for many photometric programmes even when they can accept much higher
random errors.  Eq.~(\ref{equ:zeroerr}) can be used to guide observers.  For
example, the goal to achieve a photometric zero point better than 20$\,$mmag
with 99.7\% confidence implies a $3\sigma$ error for the zero point of
20$\,$mmag.  Eq.~(\ref{equ:zeroerr}) implies that  eight standard fields are
needed. 

\subsection{Three Percent Photometry}

The above discussion shows that 3\% photometry can be reached with FORS1
with moderate effort. For the purpose of this discussion, three percent
photometry  is defined  as a total 1$\sigma$ error including both random errors
on individual star and systematic errors due to zero point. With three
calibration fields, the error in the zero point is 11$\,$mmag
(Sec.~\ref{sec:howmany}). The maximum possible systematic error is 3$\,$mmag
(Sec.~\ref{sec:err}). This leaves  $\sqrt{30\,{\rm mmag}^2-11\,{\rm
mmag}^2-3\,{\rm mmag}^2}= 27.8\,$mmag for the  possible random  error in the
magnitude of the science targets. A standard 1~hour Observing Block results in 50 minutes
of open shutter exposure time.  Using the ESO Exposure Time Calculator, we find that under standard conditions, the 3\% goal can be reached down to a $R$ band magnitude of 24.3.

\section{Conclusions}

The main result of {\em FAP} is that it is possible to achieve 3\% photometry
with FORS1 with moderate effort.  To achieve this accuracy, corrections to the
standard master flats have to be applied, and a sufficient number of standard
field observations have to be obtained. The Stetson fields seem to be well
suited for that purpose. For service observations at ESO, observations of
photometric standards  in addition to the routine nightly calibration are
charged to the science time.  Observers therefore need to consider these
calibration requirements during proposal writing and include them into the
request for observing time. The results of Sec.\ref{sec:howmany} can be used
estimate the necessary observing time for  calibration.

\begin{appendix}
\begin{center}
    {\bf APPENDIX}
\end{center}

\section{Formulae to fit $F(x,y)$ from stars without known
magnitudes}\label{sec:form}

In general, each measured magnitude on any of the  images
can be written as

\begin{equation}
m_{\nu\mu} = M_\nu + z_\mu - f(x,y)
\end{equation}

\noindent where $M_\nu$ is the magnitude of star $\nu$ within the chosen band,
$m_{\nu\mu}$ is its instrumental magnitude measured on image $\mu$, 
and $z_\mu$ is the zero point
of image $\mu$. The quantity $f(x,y)$ is $F(x,y)$ expressed in magnitudes, 
i.e. 
\begin{equation}
f(x,y)= -2.5 \log F(x,y)
\end{equation}

The specific model for $f(x,y)$  used for the fit to the current data set
is a polynomial of order $o$, 

\begin{equation}
   f(x,y)= \sum_{i=0}^o \sum_0^i p_{ij}x^iy^{o-i}. \label{fig:genequ}
\end{equation}

The magnitude  for the $n+1$ different observed stars,  $M_\nu$ where $\nu= 0
\ldots n$, and the zero points of the $m+1$ different images, $z_\mu$, $\mu= 0
\ldots m$, are further free parameters. Two of the three parameters $p_{0,0}$,
$M_0$ and $z_0$ are redundant and can be arbitrarily fixed.  Choosing $M_0 =
z_0 = 0$, the full set of equations \ref{fig:genequ} can be written as

\begin{equation}
\mathbf{A} \cdot \mathbf{p} = \mathbf{M}
\label{svdequ}
\end{equation}

\noindent where $\mathbf{p}$ and $\mathbf{M}$ are  the vectors for the parameter and  vector instrumental magnitudes, 
\begin{equation}
\mathbf{p}=
\begin{pmatrix}
     p_{0,0} \cr p_{1,0}  \cr p_{0,1} \cr \vdots \cr p_{kl} \cr M_1 \cr M_2 \cr  \vdots \cr M_n \cr  z_1 \cr z_0 \cr \vdots \cr z_m
\end{pmatrix} , \ \ 
\mathbf{M} =
\begin{pmatrix}
m_{0,0} \cr
m_{1,0} \cr
m_{2,0} \cr
\vdots \cr
m_{n,0} \cr
m_{0,1} \cr
m_{1,1} \cr
\vdots \cr
m_{n,m}
\end{pmatrix}
\end{equation}

and the matrix $\mathbf{A}$ is

\begin{equation}
\begin{array}{lll}
\mathbf{A}=  \\
\tiny
\begin{array}{rrcccccccccccccl}
    &  & p_{0,0} & p_{1,0}  & p_{0,1} & \cdots & p_{kl}        & M_1 & M_2 &  \cdots & M_n &  z_1 & z_2 & \cdots & z_m &    \cr 
0 &  
\ldelim({15}{1mm} 
&
         1 & x_{0,0}  & y_{0,0} & \cdots & x_{0,0}^ky_{0,0}^l   & 0   & 0   & \cdots & 0   & 0       & 0 & \cdots   & 0   &           
\rdelim){15}{1mm} 
\cr
1 &     & 1 & x_{1,0}  & y_{1,0} & \cdots & x_{1,0}^ky_{1,0}^l   & 1   & 0   & \cdots & 0   & 0       & 0 & \cdots   & 0   &   \cr
2 &     & 1 & x_{2,0}  & y_{2,0} & \cdots & x_{2,0}^ky_{2,0}^l   & 0   & 1   & \cdots & 0   & 0       & 0 & \cdots   & 0   &   \cr
\vdots & & & \vdots & & & & \vdots & & & & &   \vdots \cr
   n &  & 1 & x_{n,0}  & y_{n,0} & \cdots & x_{n,1}^ky_{n,1}^l   & 0   & 0   & \cdots & 1   & 0       & 0 & \cdots   & 0   &   \cr
   n+1 &  & 1 & x_{1,1}  & y_{1,1} & \cdots & x_{1,1}^ky_{1,1}^l & 0   & 0   & \cdots & 0   & 1       & 0 & \cdots   & 0   &   \cr
   n+2 &  & 1 & x_{2,1}  & y_{2,1} & \cdots & x_{2,1}^ky_{2,1}^l & 1   & 0   & \cdots & 0   & 1       & 0 & \cdots   & 0   &   \cr
\vdots & & & \vdots & & & & \vdots & & & & &   \vdots \cr
   n\times m &  & 1 & x_{n,m}  & y_{n,m} & \cdots & x_{n,m}^ky_{n,m}^l & 0   & 0   & \cdots & 1   & 0       & 0 & \cdots   & 1   &   \cr
\end{array} 
\\
\quad
\end{array}
\end{equation}

The parameters corresponding to each column are shown on the top
of the matrix.
Note that only a subset of all stars is contained in any single image, the
labelling of the rows on the left side of  the matrix is therefore not
necessarily consecutive.  The total number of free parameters to be determined
$n_p$ is
\begin{equation}
n_p= n + m + \sum_{i=0}^o(i+1) =  n + m + \frac{o^2+3o+2}{2}
\end{equation}
\noindent whereas the number of equations is identical to the number of measured
instrumental magnitudes. If the number of instrumental magnitudes per image is
$>>2$, then Eq.~(\ref{svdequ}) is an over-determined set of linear equations.

{\sl
\sindex{Singular Value Decomposition} (SVD)} can be used to find the unknown zero points,
magnitudes and model
parameters simultaneously in a least-square sense.
SVD works by decomposing the matrix $\mathbf{A}$ into a square diagonal matrix
$\mathbf{w}$ with  positive or zero elements,  and two orthogonal matrices
$\mathbf{u}$ and $\mathbf{v}$, 
\begin{equation}
\mathbf{A}= \mathbf{u}\cdot \mathbf{w} \cdot \mathbf{w^t}. 
\end{equation}
Then the least square solution for $\mathbf{M}$ can be found
as 
\begin{equation}
\mathbf{p}= \mathbf{v}\cdot \mathbf{w'}\cdot \mathbf{u^t}\cdot \mathbf{M} \label{equ:solve}
\end{equation}
where $\mathbf{w'}$ is  a matrix which consists of a the inverse of
a $n_p\times n_p$ submatrix
 of  $\mathbf{w}$ and is set to zero elsewhere \citep[see][for details]{nr}.

One consideration for solving this set of equations is to assign proper weights
to each equation.  Eq.~(\ref{svdequ}) still holds when each row in the matrix
$\mathbf{A}$ as well as corresponding elements of the vector of instrumental
magnitudes are multiplied by an arbitrary weight.  We weighted each equation
taking into account both the uncertainty in the measured instrumental magnitudes and
the local density of stars.

The estimated uncertainty $\sigma_{\nu\mu}$ in the instrumental
magnitude of the $\nu^{\it th}$ star in the $\mu^{\it th}$ field as
given by Sextractor were used to compute a weight $w_m$, 

\begin{equation}
   w_m= \frac{1}{\sigma_{\nu\mu}^2}
\end{equation}

A significant source of uncertainty in the fit of our model to the zero points
is the difference between the true shape of $f(x,y)$ and that of the model
polynomial. If an unweighted fit of a polynomial  were used, more weight would
be given  to regions with high density of observed stars. This would introduce
biases in the fit which can be avoided by adjusting the weights according to
the local density of stars.   Specifically, we have used  the inverse of the
local density of $w_m$ to compute a second weight $w_\rho$, 
\begin{equation}
   w_\rho= \frac{1 }{\sum w_m}
\end{equation}
where the sum in this equation is taken over all magnitude measurements
in cells of $128\times 128$ pixels on the detector. The final weight
used for each equation was
\begin{equation}
   w_t= w_\rho \cdot w_m
\end{equation}

\section{Formulae to fit $F(x,y)$ and extinction solution simultaneously}\label{sec:form_mag}

The formulae in Appendix~\ref{sec:form} can easily be modified when the magnitudes of stars are known.
The magnitude zero points for individual stars are replaced with the colour
term, the zero points for individual images are replaced by the extinction
terms, and the zero of the polynomial $p_{0,0}$ is replace by the constant 
magnitude zero point to find the parameters of Eq.~(\ref{equ:ext}).

The parameter vector $\mathbf{p}$ and 
$\mathbf{M}$ are in this case 
\begin{equation}
\mathbf{p}=
\begin{pmatrix}
     z \cr p_{1,0}  \cr p_{0,1} \cr \vdots \cr p_{kl} \cr e \cr  a \cr  c 
\end{pmatrix} 
\rm{,\ \ \ }
\mathbf{M} =
\begin{pmatrix}
M_{0}-m_{0,0} \cr
M_{1}-m_{1,0} \cr
M_{2}-m_{2,0} \cr
\vdots \cr
M_{n}-m_{n,0} \cr
M_{0}-m_{0,1} \cr
M_{1}-m_{1,1} \cr
\vdots \cr
M_{n}-m_{n,m}
\end{pmatrix}
\end{equation}

\noindent and the matrix $\mathbf{A}$ is

\begin{equation}
\begin{array}{lll}
\mathbf{A}=  \\
\begin{array}{rrcccccccccccccl}
    &  & z & p_{1,0}  & p_{0,1} & \cdots & p_{kl}        & e & a &  c    \cr 
0 &  
\ldelim({10}{1mm} 
&
         1 & x_{0,0}  & y_{0,0} & \cdots & x_{0,0}^ky_{0,0}^l   & X_0   & c_0   &  X_0\times c_0      &           
\rdelim){10}{1mm} 
\cr
1 &     & 1 & x_{1,0}  & y_{1,0} & \cdots & x_{1,0}^ky_{1,0}^l   & X_0  & c_1   &  X_0\times c_1   &   \cr
2 &     & 1 & x_{2,0}  & y_{2,0} & \cdots & x_{2,0}^ky_{2,0}^l   & X_0  & c_2   &  X_0\times c_2   &   \cr
\vdots & & & \vdots & & & & \vdots  \cr
   n &  & 1 & x_{n,0}  & y_{n,0} & \cdots & x_{n,1}^ky_{n,1}^l   & X_1  & c_n   &  X_1\times c_n   &   \cr
   n+1 &  & 1 & x_{1,1}  & y_{1,1} & \cdots & x_{1,1}^ky_{1,1}^l & X_1  & c_1   &  X_1\times c_1   &   \cr
   n+2 &  & 1 & x_{2,1}  & y_{2,1} & \cdots & x_{2,1}^ky_{2,1}^l & X_1  & c_2   &  X_1\times c_2   &   \cr
\vdots & & & \vdots & & & & \vdots  \cr
   n\times m &  & 1 & x_{n,m}  & y_{n,m} & \cdots & x_{n,m}^ky_{n,m}^l & X_m & c_n   &   X_m\times c_n   &   \cr
\end{array} 
\\
\quad
\end{array}
\end{equation}

where $X_\mu$ is the airmass of the $\nu$th image, and $c_\nu$ is the R-I colour of the $\nu$th star.
The least square solution for this set of linear equations can be found as above using
Eq.~\ref{equ:solve}.

\end{appendix}

\acknowledgements We thank Jason Spyromilio for granting technical time on
FORS1 to carry out the {\em FAP} observations, David Silva and Sabine Moehler
for useful discussions, and Jeremy Walsh for proof reading an earlier version
of the  {\em FAP} report.

\end{document}